\documentclass[fleqn,usenatbib]{mnras}

\usepackage{newtxtext,newtxmath}

\usepackage[T1]{fontenc}

\DeclareRobustCommand{\VAN}[3]{#2}
\let\VANthebibliography\thebibliography
\def\thebibliography{\DeclareRobustCommand{\VAN}[3]{##3}\VANthebibliography}

\usepackage{graphicx}	
\usepackage{amsmath}	

\newcommand{\vek}[1]{\mathbf{#1}}
\newcommand{\pdv}[2]{\frac{\partial #1}{\partial #2}}

\newcommand{\rot}{\mathbf{\nabla}\times}

\newcommand{\Tfrac}[2]{\left(\frac{#1}{#2}\right)}

\newcommand{\vgrad}[1]{\mathbf{\nabla}{#1}}

\newcommand*\diff{\mathop{}\!\mathrm{d}}

\DeclareMathAlphabet\mathbfcal{OMS}{cmsy}{b}{n}

\usepackage{lipsum}

\newcommand{\vEMF}{\boldsymbol{{\mathcal{E}}}}
\newcommand{\mean}[1]{\left\langle #1 \right\rangle}

\newcommand{\vmean}[1]{\left\langle \mathbf{#1} \right\rangle}

\title[]{Thinner Disks Struggle to Launch Jets: GRMHD Simulations of Magnetic Field Generation and Transport}

\author[Jonatan Jacquemin-Ide et al.]{
Jonatan Jacquemin-Ide,$^{1}$\thanks{E-mail: Jonatan.Jacquemin@colorado.edu}
Mitchell C. Begelman,$^{1,2}$
Henrik Latter,$^{3}$
Jason Dexter,$^{1,2}$\newauthor
and Alexander Tchekhovskoy$^{4}$
\\
$^{1}$JILA, University of Colorado and National Institute of Standards and Technology, 440 UCB, Boulder, CO 80309-0440, USA\\
$^{2}$Department of Astrophysical and Planetary Sciences, University of Colorado, 391 UCB, Boulder, CO 80309-0391, USA\\
$^{3}$Department of Applied Mathematics and Theoretical Physics, University of Cambridge, Wilberforce Road, Cambridge CB3 0WA, UK\\
$^{4}$Center for Interdisciplinary Exploration $\&$ Research in Astrophysics (CIERA), Physics and Astronomy, Northwestern University, Evanston, IL 60202, USA
}

\date{Accepted XXX. Received YYY; in original form ZZZ}

\pubyear{\the\year{}}

\begin{document}
\label{firstpage}
\pagerange{\pageref{firstpage}--\pageref{lastpage}}
\maketitle

\begin{abstract}
Relativistic jets are observed in some accreting black-hole systems but not others. In X-ray binaries, jet production is closely associated with geometrically thicker accretion states, suggesting that disk thickness may influence the conditions for jet launching. We investigate this connection using three-dimensional GRMHD simulations of disks with $h/r=0.1$--$0.3$, initialized with a toroidal magnetic field with plasma $\beta=5$.

We find that the magnetorotational instability (MRI) generates large-scale poloidal fields  through non-axisymmetric shearing waves, with an additional contribution from axisymmetric induction through stretching by channel-like modes. The resulting mean fields become progressively weaker as the disk becomes thinner. Because these fields regulate MRI turbulence, thinner disks exhibit weaker angular-momentum transport and accrete more slowly than the standard $v_{\rm acc}\propto(h/r)^2$ thin-disk scaling would predict. The magnetic-flux transport velocity remains comparable to the accretion velocity, so weaker field generation and slower transport combine to strongly suppress black-hole flux accumulation at smaller $h/r$. Correspondingly, the thinnest disk ($h/r=0.1$) produces an electromagnetic jet efficiency of only $\eta_{\rm{EM}}\sim1\%$, while the thickest disk reaches the magnetically arrested disk (MAD) state and attains $\eta_{\rm{EM}}\geq100\%$.

Using the measured large-scale magnetic flux distribution and transport velocities, we estimate the time required to accumulate MAD-level magnetic flux. Only the $h/r\sim0.3$ disk reaches MAD in the duration of our simulations, at $t\simeq2\times10^4\,r_g/c$, whereas for $h/r=0.1$ we estimate a timescale of $\sim10^8\,r_g/c$. These results link disk geometry to magnetic-flux accumulation and jet launching, with direct implications for the timescales over which jets can emerge in transient accreting systems.
\end{abstract}

\begin{keywords}
accretion, accretion discs – black hole physics – magnetic fields – MHD – jets
\end{keywords}



\section{Introduction}
\label{sec:intro}

Relativistic jets provide a powerful channel through which accreting black holes release energy and angular momentum, from stellar-mass black holes to supermassive black holes (SMBHs). They accompany some of the most luminous transient events, including gamma-ray bursts (GRBs), whose central engines can arise from compact-object mergers or the collapse of massive stars \citep{woosley_gamma-ray_1993,metzger_kilonovae_2020,rastinejad_kilonova_2022}. A subset of tidal disruption events (TDEs) also produces powerful relativistic jets \citep{zauderer_birth_2011,burrows_relativistic_2011,cenko_swift_2012,pasham_birth_2023}, while jets from active galactic nuclei (AGN) can inject substantial energy into their host galaxies and larger-scale environments \citep{blandford_relativistic_2019,hardcastle_radio_2020}. Relativistic jets can extract rotational energy from the black hole and drive its spin evolution toward lower values \citep{lowell_rapid_2024,lowell_evidence_2025}. Understanding why some accreting black holes launch powerful jets while others do not is therefore a fundamental problem across a wide range of astrophysical systems.

Observations demonstrate that jet production is intimately connected to the state of the accretion flow, but that accretion power alone does not determine whether a jet is present. In X-ray binaries (XRBs), persistent compact jets are associated with the hard state, in which the inner flow is hot and geometrically thick, whereas jets are strongly quenched when the system transitions to the soft, thin-disk state \citep{done_modelling_2007,fender_towards_2004}. Powerful transient ejecta can also accompany these state transitions \citep{fender_towards_2004}. A similar connection between accretion geometry and jet production may extend to AGN, where radio-loud systems are preferentially associated with radiatively inefficient, geometrically thick accretion flows \citep{narayan_advection-dominated_1994,meier_association_2001}, although only a fraction of AGN are strongly jetted \citep{padovani_microjansky_2011}. The connection between inflow and outflow is also reflected in the observed radio--X-ray correlation in hard-state XRBs \citep{corbel_radio/x-ray_2003,gallo_universal_2003,corbel_universal_2013} and, after accounting for black-hole mass, in the fundamental plane extending from stellar-mass black holes to AGN \citep{merloni_fundamental_2003,zamaninasab_dynamically_2014}.

TDEs provide another suggestive example. Their accretion flows are expected to evolve from initially super-Eddington, geometrically thick configurations toward thinner disks as the fallback and accretion rates decline \citep{strubbe_optical_2009,dai_unified_2018}. While only a small subset are observed to produce prompt relativistic jets \citep{zauderer_birth_2011,burrows_relativistic_2011,tchekhovskoy_swift_2014,pasham_birth_2023}, late-time radio observations have revealed outflows launched hundreds to thousands of days after disruption in a broader population of TDEs \citep{cendes_mildly_2022,cendes_ubiquitous_2024}. Together, these observations suggest that the structure and geometry of the accretion flow may play a fundamental role in regulating both the formation and timing of relativistic outflows.

Magnetic flux provides a natural link between accretion and jet launching. Large-scale magnetic flux threading the inner accretion flow and black-hole horizon is required to efficiently extract energy and angular momentum and power a relativistic jet \citep{blandford_electromagnetic_1977,blandford_hydromagnetic_1982}. At the same time, magnetic fields are essential for accretion itself, either by driving turbulence through the magnetorotational instability (MRI) \citep{balbus_powerful_1991,balbus_instability_1998} or by exerting large-scale magnetic torques on the accretion disk \citep{blandford_hydromagnetic_1982,ferreira_magnetized_1995}.

Thus, if jet launching requires a large-scale organized magnetic field, the accretion flow must not only release gravitational energy but also generate, organize, and transport magnetic flux toward the black hole. The key question is therefore whether the field can form coherent large-scale structures and be advected inward against turbulent diffusion \citep{lubow_magnetic_1994,guilet_transport_2012,begelman_simple_2024}. This question might be related to why systems with similar accretion luminosities can produce very different jet powers, and why individual systems can transition between jetted and unjetted states.

This problem is particularly important because accretion flows do not naturally begin with the coherent vertical magnetic flux required for jet launching, which is commonly imposed in numerical simulations. In dynamically assembled disks, strong shear preferentially amplifies azimuthal magnetic fields, as expected in tidal disruption events and black-hole–neutron-star (BHNS) and binary-neutron-star (BNS) mergers \citep{bonnerot_magnetic_2017,most_accretion_2021,aguilera-miret_role_2023}. A similar situation arises during AGN disk formation, where the accretion of magnetized intergalactic or interstellar plasma is expected to produce predominantly azimuthal fields rather than coherent vertical flux \citep{gaburov_magnetically_2012,hopkins_forged_2024,squire_rapid_2025,kaaz_h-amr_2025}.
Finally, collapsar events also show an initial toroidal field present after the accretion disk is formed \citep{chan_jets_2026}. 
Moreover, the progenitors of TDEs and compact-object mergers are not generally expected to supply enough coherent large-scale magnetic flux to account directly for the most powerful observed jets, such as that of Swift J1644+57 \citep{tchekhovskoy_swift_2014}. These systems therefore point toward a common requirement: a mechanism must generate and organize large-scale poloidal magnetic flux from fields that are initially predominantly small-scale and toroidal.

Understanding this process is particularly timely given the growing observational access to horizon-scale magnetized plasma through GRAVITY and the Event Horizon Telescope (EHT) \citep{jimenez-rosales_dynamically_2020,collaboration_first_2021}. A physical description of magnetic-field generation and transport can therefore provide the missing link between the structure of accretion flows and the emergence, strength, and intermittency of relativistic jets.

Perhaps the key insight necessary to make progress is that the MRI does not just drive accretion of material, it also acts as a dynamo, providing a route to build large-scale poloidal magnetic fields from the predominantly toroidal fields naturally produced in accretion disks. This possibility was recognized shortly after the discovery of the MRI \citep{brandenburg_dynamo-generated_1995,hawley_local_1995,hawley_local_1996}. In particular, MRI turbulence does not require a net vertical magnetic flux to persist: even from zero-net-flux initial conditions, the instability can regenerate the magnetic field needed to sustain itself \citep{rincon_self-sustaining_2007}. The associated angular momentum transport remains relatively weak \citep{shakura_black_1973,fromang_mhd_2007-1}.

The self-sustained state results from a nonlinear cycle that has been extensively investigated using non-stratified shearing-box simulations \citep{lesur_self-sustained_2008,herault_periodic_2011,riols_global_2013,riols_dissipative_2015,riols_magnetorotational_2017,mamatsashvili_zero_2020,held_mri_2022}. In this picture, Keplerian shear converts a weak, axisymmetric poloidal field into a toroidal component. The latter can then become unstable to non-axisymmetric MRI perturbations, which take the form of MHD shearing waves \citep{goldreich_ii_1965,balbus_powerful_1992,johnson_magnetohydrodynamic_2007}. Their nonlinear interactions regenerate the large-scale poloidal field, allowing shear to act on it again and complete the cycle. Similar results have recently been found in stratified shearing boxes\footnote{albeit with some dissimilarities due to the presence of Parker modes} \citep{held_mri_2024}.
Because of the nonlinear stage of evolution, this process is fundamentally different from a kinematic dynamo and requires a finite magnetic seed to become established \citep{rincon_dynamo_2019}.

The presence of vertical stratification introduces a characteristic temporal and spatial structure to this process. The large-scale field undergoes quasi-periodic polarity reversals that propagate from the disk midplane toward the corona, producing the familiar butterfly pattern \citep{brandenburg_dynamo-generated_1995,davis_sustained_2010,gressel_mean-field_2010,gressel_characterizing_2015,salvesen_accretion_2016}. 
These cycles have become a common feature of weakly magnetized global simulations of accretion disks \citep{flock_turbulence_2011,flock_large-scale_2012,hogg_influence_2018,roberts_global_2025}.
Similar propagating field reversals have also been reported in global GRMHD simulations of black-hole--neutron-star mergers \citep{hayashi_general-relativistic_2022,hayashi_general-relativistic_2023}, indicating that MRI dynamo action may operate in self-consistently formed accretion disks and potentially provide a mechanism for generating large-scale magnetic fields without imposing them as an initial condition.

A purely toroidal magnetic field can nevertheless evolve into a magnetically arrested disk (MAD), even though the MAD state is generally understood as the saturated configuration in which the black hole has accumulated the maximum sustainable amount of large-scale poloidal magnetic flux \citep{tchekhovskoy_efficient_2011,liska_large-scale_2020}. The key ingredient is the MRI dynamo, which generates large-scale poloidal magnetic loops that are subsequently advected inward through the disk \citep{jacquemin-ide_magnetorotational_2024}. As the simulation evolves, loops originating at progressively larger radii can reach the black hole. Because the local thermal scale height increases with radius, these loops also have larger characteristic scales and can carry more coherent poloidal magnetic flux. The inward transport of progressively larger magnetic structures can therefore build up the black-hole flux until the system eventually reaches the MAD state.

In this paper, we extend the results of \cite{jacquemin-ide_magnetorotational_2024} to accretion disks with different thicknesses, focusing on how disk geometry affects the generation and transport of large-scale magnetic fields. We initialize our simulations with a mean toroidal magnetic field and no large-scale poloidal field, an initial configuration expected in dynamically assembled accretion flows such as those formed in TDEs, mergers, and collapsars \citep{gaburov_magnetically_2012,bonnerot_magnetic_2017,hopkins_forged_2024,most_accretion_2021,aguilera-miret_role_2023,kaaz_h-amr_2025,chan_jets_2026}.

We find that the strength of MRI-driven turbulence depends strongly on the disk thickness, rather than simply on the initial toroidal field strength. The magnetic flux accumulated on the black hole also decreases steeply with decreasing $h/r$. We trace this dependence to the combined effect of disk thickness on the strength of the generated poloidal field and on the rate at which this field is transported inward. \emph{These results provide a first-principles explanation for the connection between disk thickness and jet launching, and bring us closer to understanding the physical conditions required for the formation of jets.}

In Section~\ref{sec:num_meth}, we describe the numerical code, initial conditions, and analysis methodology. Section~\ref{sec:descript} presents the main features of our simulations, including the accumulation of magnetic flux (Section~\ref{sec:flux_hole}), the emergence of poloidal magnetic field lines and magnetic elevation (Section~\ref{sec:glob_stru}), and the local turbulent structure (Section~\ref{sec:turbulent_behavior}). We then investigate the generation of the poloidal and toroidal fields in detail in Section~\ref{sec:field_gen}, with the goal of understanding why thinner disks are less efficient at generating large-scale poloidal magnetic fields. In Section~\ref{sec:transport}, we examine the transport of angular momentum (Section~\ref{sec:ang_transport}) and magnetic fields (Section~\ref{sec:field_transport}). In Section~\ref{sec:implications}, we compile the scalings of field generation and transport with $h/r$ (Section~\ref{sec:scalings}) and formulate a simple predictions for the rate at which magnetic flux accumulates on the black hole (Section~\ref{sec:time_to_MAD}). Finally, in Section~\ref{sec:obs_cons}, we discuss the observational consequences of our results, while Section~\ref{sec:context_limits} discusses their theoretical context and numerical limitations. We summarize our conclusions in Section~\ref{sec:conclu}.

\section{Methods}
\label{sec:num_meth}
\begin{table*}
    \centering
    \caption{Properties of the simulations considered in this work. The averaging times are far earlier for the thickest simulation because this simulation reaches the MAD state at $t\sim10^4$.}
    \label{tab:simulations}
    \scriptsize
    \setlength{\tabcolsep}{3pt}
    \begin{tabular}{lcccccc}
        \hline
        $h/r$ &  Effective Resolution $N_R\times N_\theta\times N_\varphi$ & Averaging interval in $r_g/c$ & Averaging interval in local shear timescale, $\Omega_k^{-1}$ at $r=20r_g$ & Dumping interval & Number of dumps  \\
        \hline
        $\sim 0.3$ & $1872\times624\times1024$ & $t_1 =4\times10^3$ to $t_2 = 9\times10^3$ & $t_1 =44$ to $t_2 =100$ & $\Delta t = 10\,\,r_g/c=0.11\Omega_k^{-1}|_{r=20r_g}$ & 500 dumps \\
        $0.2$      & $5376\times2304\times4096$ & $t_1 =3\times10^4$ to $t_2 = 4\times10^4$ & $t_1 =335$ to $t_2 =447$ & $\Delta t = 50\,\,r_g/c=0.55\Omega_k^{-1}|_{r=20r_g}$ & 200 dumps  \\
        $0.1$      & $5376\times2304\times4096$ & $t_1 =3\times10^4$ to $t_2 = 4\times10^4$ & $t_1 =335$ to $t_2 =447$ & $\Delta t = 50\,\,r_g/c=0.55\Omega_k^{-1}|_{r=20r_g}$ & 200 dumps \\
        \hline
    \end{tabular}
\end{table*}
\subsection{Numerical methods}

We perform our simulations with the \texttt{H-AMR} code \citep{Liska2022}, which integrates the ideal general-relativistic magnetohydrodynamic (GRMHD) equations on a spherical-polar $(r,\theta,\varphi)$ mesh using Kerr-Schild coordinates \citep{gammie_harm_2003}. For reference, we additionally introduce the cylindrical coordinates $R$ and $z$ to describe the radial and vertical structure of the flow. Throughout the simulations, we work in geometrized units, setting $G=M=c=1$, with $M$ denoting the black hole mass; consequently, the gravitational radius $r_{\rm g}=GM/c^2$ is also unity. The magnetic field is expressed in Lorentz-Heaviside units, for which the magnetic pressure in the fluid frame is $p_{\rm mag}=b^2/2$, where $b$ is the magnitude of the magnetic-field four-vector measured in the fluid frame. Spatial reconstruction is performed using the piecewise parabolic method \citep{Colella}, while the temporal integration employs a second-order scheme. We use an HLL scheme for the Riemann solver.

We perform three GRMHD simulations with different disk thicknesses. The thickest case is the simulation analyzed by \cite{jacquemin-ide_magnetorotational_2024} and originally presented in \cite{liska_large-scale_2020}. It follows a torus orbiting a rapidly rotating black hole with $a=0.9$, using a ideal-gas equation of state with polytropic index $\gamma=5/3$. The torus extends from $r_{\rm in}=6\,r_g$ to $r_{\rm out}\approx4\times10^4\,r_g$, with a pressure maximum at $r_{\rm max}=13.792\,r_g$. Although its initial thickness is $h/r\sim0.2$ near the pressure maximum, the adiabatic evolution leads to a disk with $h/r\sim0.3$, where $h$ is the thermal scale height. 
The initial magnetic field is purely toroidal and corresponds to $\beta_{\rm ini}=5$. The simulation is evolved with \texttt{H-AMR} at an effective resolution of $1872\times624\times1024$ over most of the disk, corresponding to approximately 70--90 cells per scale height. Further details of the initial conditions and numerical setup can be found in \cite{liska_large-scale_2020,jacquemin-ide_magnetorotational_2024}.

Our two thinner-disk simulations instead start from a prescribed disk structure, with a Gaussian density profile in the latitudinal direction and a radial density profile $\rho\propto r^{-1}$. The black hole spin for both of these simulations is set to $a=0.9375$. They have the initial thicknesses $h/r=0.1$ and $0.2$, and extend from $r_{\rm in}=6.5,r_g$ to $r_{\rm out}=3000\,r_g$, while the computational grid reaches out to $r=10^4\,r_g$. We use the cooling prescription of \cite{noble_direct_2009} to maintain their respective disk thicknesses, and initialize both models with a purely toroidal magnetic field with $\beta_{\rm ini}=5$. The base resolution\footnote{This refers to the resolution near the disk midplane; external derefinement is used to reduce the $\phi$-resolution toward the poles to alleviate the CFL constraint \citep{Liska2022}.} is $1344\times576\times1024$. Two additional levels of static mesh refinement are applied around the disk midplane at $|z|/R=0.268$ and $|z|/R=0.132$, increasing the effective resolution there to $5376\times2304\times4096$. This corresponds to 64 cells per scale height for the $h/r=0.1$ model and 128 cells per scale height for the $h/r=0.2$ model. In Appendix~\ref{A:resol}, we check our resolution against the commonly used standard quality factors \citep{hawley_assessing_2011,hawley_testing_2013} and the newly proposed criteria of \citet{jannaud_numerical_2026}. We find that the simulations are mostly resolved according to these criteria.

\subsection{Analysis methods}

We separate large-scale/average and fluctuating components using a Reynolds decomposition,
\begin{equation}
X=\mean{X}+\delta X,
\label{Eq:Rey_avg}
\end{equation}
with the averaging operator defined below. By construction, the fluctuations satisfy $\mean{\delta X}=0$, whereas products of fluctuations can retain a non-zero mean. Terms such as $\mean{\delta X\delta Y}$ therefore describe the nonlinear influence of the turbulent fluctuations on the large-scale flow.

Here, the averaging operation is performed over the azimuthal direction,
\begin{equation}
\mean{X}=\frac{1}{2\pi}\int_0^{2\pi}X\,\rm{d}\varphi.
\end{equation}
We consequently refer to $\mean{X}$ as either the mean or axisymmetric component. In addition to this azimuthal averaging, we occasionally characterize quantities over a finite latitudinal extent of the disk. We define this disk average as
\begin{equation}
X|_{d}=
\frac{1}{\int_{\theta_1}^{\theta_2}\sqrt{-g}\,\rm{d}\theta}
\int_{\theta_1}^{\theta_2}\sqrt{-g}\,\mean{X}\,\rm{d}\theta,
\label{Eq:disk_avg}
\end{equation}
where
\begin{equation}
\theta_{1,2}=\frac{\pi}{2}\pm\arctan\left(N_h\frac{h}{R}\right).
\end{equation}
This operation is applied only after azimuthal averaging. Depending on the simulation, $N_h$ is chosen so that the interval encompasses the turbulent disk while excluding the jet funnel.

For the magnetic-field decomposition, we distinguish the poloidal component,
\begin{equation}
\mathbf{B_p}=B_r\vek{e}_r+B_\theta\vek{e}_\theta.
\end{equation}
We use physical components throughout unless explicitly stated otherwise, so that the polar component in this expression corresponds to $B{\hat{\theta}}=\sqrt{g_{\theta\theta}}B^\theta$. Hats are suppressed hereafter for notational convenience.

Our analysis of magnetic-field generation is based on the induction equation. At the radii of interest, since the field is predominantly generated far from the black hole, we can use its Newtonian form,
\begin{equation}
\pdv{\mathbf{B}}{t}
=\vgrad{}\times\left(\mathbf{u}\times\mathbf{B}\right).
\label{Eq:Newt_induc}
\end{equation}
We assess the validity of this approximation a posteriori by comparing the two sides of Eq.~(\ref{Eq:Newt_induc}); good agreement indicates that the Newtonian treatment is adequate at the radii considered and beyond.

Applying the azimuthal average to Eq.~(\ref{Eq:Newt_induc}) gives
\begin{equation}
\pdv{\vmean{B}}{t}
=\mathbf{\nabla}\times
\left(\vmean{u}\times\vmean{B}+\vEMF\right),
\label{Eq:avg_induct}
\end{equation}
where nabla only contains derivatives in  $r$ and $\theta$. The additional term,
\begin{equation}
\vEMF=\mean{\delta\mathbf{u}\times\delta\mathbf{B}},
\end{equation}
is the turbulent electromotive force. It contains the contribution of correlated fluctuations to the evolution of the mean magnetic field and can either amplify or dissipate it. In particular, the EMF provides the nonlinear coupling between the mean poloidal and toroidal fields that allows the magnetic dynamo to operate and sustain itself.

We also define the shear rate and  as
\begin{equation}
    \Omega_k = \frac{1}{a+R^{3/2}},
\end{equation}
and Keplerian velocity as $V_k=R\Omega_k$.

\section{Description of the simulations}
\label{sec:descript}

In this section, we describe the evolution of our three simulations, focusing on how disk thickness affects magnetic flux accumulation and jet launching. We then examine the resulting large-scale magnetic-field geometry and the evolution of MRI turbulence.
\begin{figure}
	\includegraphics[width=\linewidth]{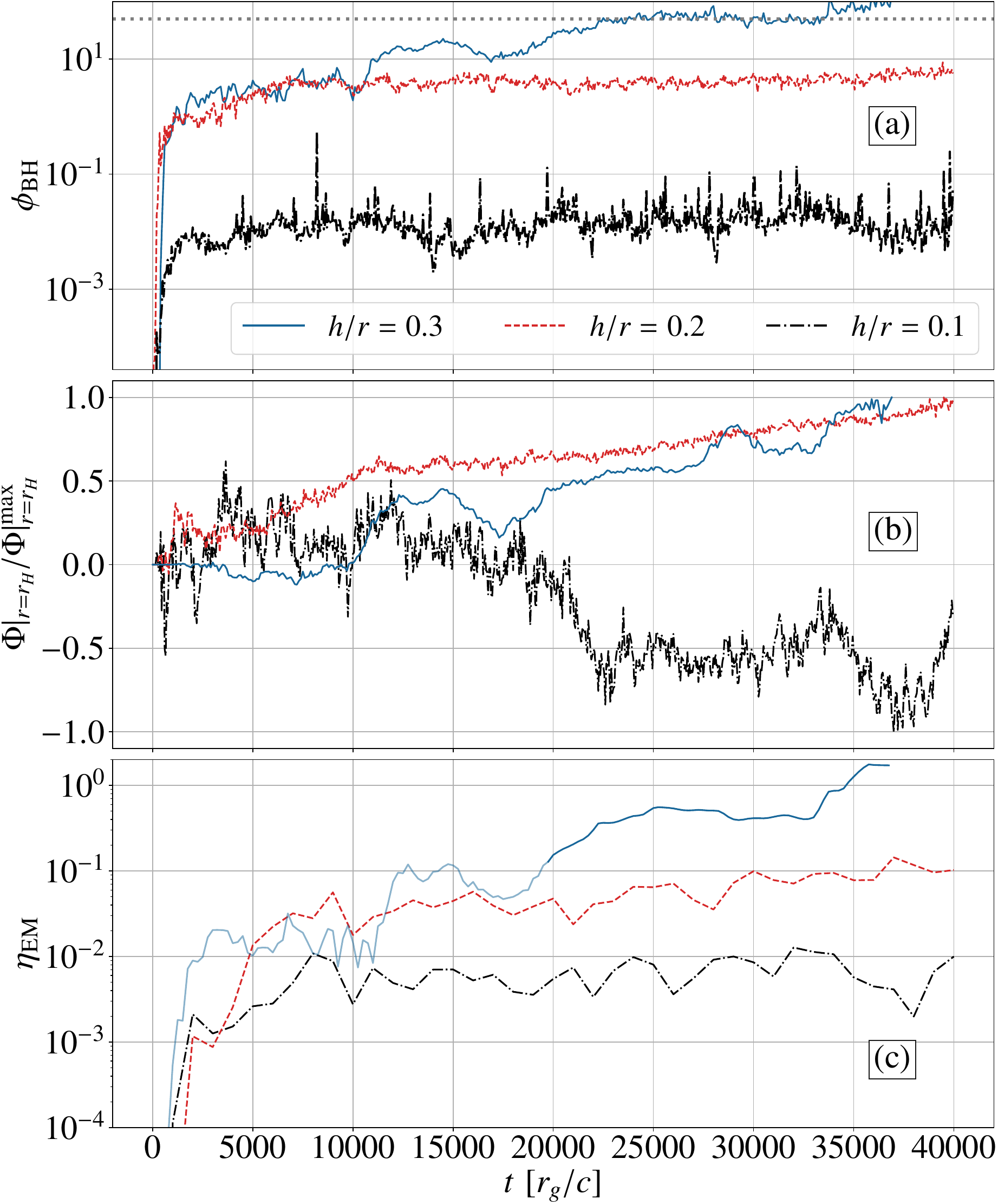}
    \caption{Black-hole magnetic flux and jet evolution for the three disk thicknesses. Panel (a) shows the MADness parameter, $\phi_{\rm BH}$, with only the thickest disk reaching the MAD state during the simulation. Panel (b) shows the signed horizon-threading magnetic flux, normalized by its maximum absolute value in each simulation, with all three cases undergoing at least one polarity reversal. Panel (c) shows the electromagnetic jet efficiency, $\eta_{\rm EM}$, measured at the black-hole horizon. For the $h/r\sim0.3$ simulation, regions shown in lighter colours indicate intervals where the data were reconstructed from the available reduced quantities. Both magnetic-flux accumulation and jet efficiency decrease strongly with decreasing disk thickness.}
    \label{fig:phi_BH}
\end{figure} 
\begin{figure*}
	\includegraphics[width=0.7\textwidth]{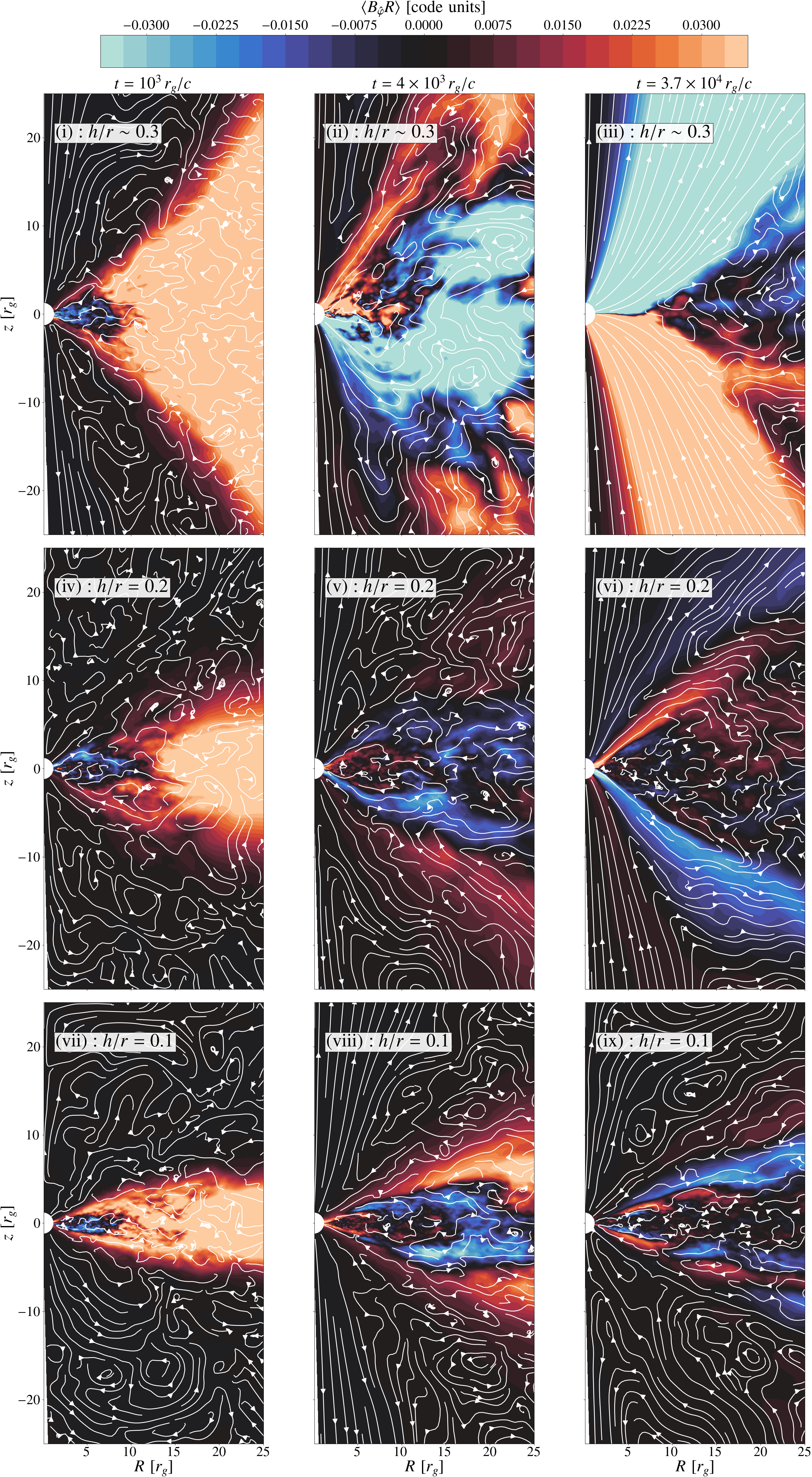}
    \caption{Poloidal slices of the three simulations showing the $\varphi$-averaged toroidal magnetic field in colour and the $\varphi$-averaged poloidal field lines as white streamlines for $r\leq25r_g$. Columns show $t=10^3$, $4\times10^3$, and $3.7\times10^4\,r_g/c$, while rows correspond to $h/r\sim0.3$, $0.2$, and $0.1$. The magnetic field becomes more organized with time, with the degree of large-scale coherence decreasing strongly toward thinner disks.}
    \label{fig:field_lines}
\end{figure*} 
\subsection{Flux on the black hole}
\label{sec:flux_hole}

We first compute the magnetic flux contained within the accretion disk, 
\begin{equation}
\label{eq:abs_phi}
    \phi(r,t) = \frac{1}{\sqrt{\dot{m}|_{5r_g} }} 2\pi \left(\int\limits_0^{\pi/2}|\mean{ B^{r}}| \sqrt{-g}\diff \theta\right),
\end{equation}
where $B^r$ is the contravariant component of the radial magnetic field and we normalize as in  \cite{tchekhovskoy_efficient_2011}. We use an absolute value to average over changes in polarity within the accretion disk, in constrast to the signed flux shown below.  We also define the accretion rate, here measured at $5\,r_g$ to avoid contamination by the numerical floors,
\begin{equation}
\label{Eq:acc_rate}
    \dot{m}|_{5r_g} = -2\pi \int\limits_0^{\pi/2} \mean{\rho u^r} \sqrt{-g}\diff \theta,
\end{equation}
where $ u^r$ is the contravariant component of the radial velocity. 
We then define the flux on the BH, also known as the MADness parameter,
\begin{equation}
\label{eq:phi_BH}
    \phi_{\rm BH} = \phi(r=r_\mathrm{H},t),
\end{equation}
where $r_\mathrm{H}$ is the event horizon.

In Fig.~\ref{fig:phi_BH}(a), we show the MADness parameter, $\phi_{\rm BH}$, as a function of time for our three simulations. In all cases, the simulations begin with zero magnetic flux, which is subsequently generated and advected onto the black hole, causing $\phi_{\rm BH}$ to increase with time. However, the growth rate differs significantly between simulations. The thick disk ($h/r\sim0.3$) reaches the MAD state, conventionally defined by $\phi_{\rm BH}=50$, by $t=2\times10^{4}\,r_g/c$. In contrast, the thinner disks do not approach this state during the duration of our simulations, which extend to $t=4\times10^{4}\,r_g/c$. At $t=2\times10^{4}\,r_g/c$, the $h/r=0.2$ disk reaches only $\phi_{\rm BH}\sim5$, while the thinnest disk ($h/r=0.1$) remains several orders of magnitude lower, at $\phi_{\rm BH}\sim3\times10^{-2}$. 

The differences between the simulations may appear dramatic and difficult to explain. However, they arise naturally from the combination of several steep scalings with disk thickness. As shown below, the strength and characteristic size of the mean magnetic field, as well as the strength of the turbulence, all depend strongly on $h/r$. These scalings in turn determine the magnetic-flux advection and accretion velocities, which also decrease rapidly with decreasing $h/r$. The combined effect of these dependencies leads to a very steep scaling of the magnetic flux threading the black hole. 

In Fig.~\ref{fig:phi_BH}(b), we show the signed magnetic flux $\Phi$ threading the black hole, 
\begin{equation}
    \Phi(t)|_{r=r_H} = 2\pi \left(\int\limits_0^{\pi/2}\mean{ B^{r}}|_{r=r_H} \sqrt{-g}\,\mathrm{d}\theta\right),
    \label{Eq:pol_flux_rg}
\end{equation}
computed over the northern hemisphere and normalized by its maximum absolute value observed during the full evolution of that simulation, $\max\|\Phi(t)|_{r=r_H}\|$. The flux is evaluated at the event horizon.  In contrast to the absolute flux shown in Fig.~\ref{fig:phi_BH}(a), the signed flux changes polarity at least once during its evolution in all three simulations. At later times, the flux appears to settle toward a single polarity, although there is no guarantee that it will not eventually reverse again.

Polarity reversals are a natural feature of dynamo-driven mean-field generation, as different magnetic polarities are generated in the disk and are subsequently advected toward the black hole \citep{christie_role_2019,jacquemin-ide_magnetorotational_2024,kaaz_h-amr_2025,chan_jets_2026}. These reversals are particularly interesting because the interaction between oppositely polarized magnetic structures may provide sites for magnetic reconnection and particle acceleration, potentially contributing to high-energy emission \citep{lyubarsky_particle_2008,sironi_acceleration_2011}. In Section~\ref{sec:transport}, we analyze the transport of magnetic loops with different polarities toward the black hole. They might also explain non-steady jet emission observed in transients see discussion in section \ref{sec:obs_cons}.

{In Fig.~\ref{fig:phi_BH}(c), we show the electromagnetic jet efficiency,}
\begin{equation}
\eta_{\rm EM}
=
\frac{-\displaystyle\int
\left(b^2u^r u_t-b^r b_t\right)
\sqrt{-g}\mathrm{d}\theta\mathrm{d}\varphi
\big|_{r=r_H}}
{\dot{m}|_{5r_g}c^2},
\end{equation}
{where $b^\mu$ is the magnetic four-vector measured in the fluid frame and the integral is evaluated over the black-hole horizon. For the thickest-disk simulation, the full output at $t\lesssim2\times10^4\,r_g/c$ was lost, so we reconstruct the jet power over this interval from the available azimuthally averaged mean-field quantities (shown as lighter blue). This reconstruction is used only for visual continuity and does not affect our conclusions, which rely on the late-time saturation of $\eta_{\rm EM}$. The full data can be found in \cite{liska_large-scale_2020}. Other diagnostics in this work use reduced data products computed before the loss of the full output.}

{For $t\gtrsim2\times10^4\,r_g/c$, the electromagnetic jet efficiency shows a steep dependence on disk thickness. The thickest disk reaches the MAD state and attains $\eta_{\rm EM}\sim2$, consistent with efficient extraction of black-hole rotational energy in MAD systems \citep{tchekhovskoy_efficient_2011,liska_large-scale_2020}. In contrast, the two thinner disks reach much smaller efficiencies, $\eta_{\rm EM}\sim0.1$ for $h/r=0.2$ and $\eta_{\rm EM}\sim0.01$ for $h/r=0.1$, corresponding to roughly an order-of-magnitude decrease for each step toward smaller disk thickness. This strong dependence may have important implications for jet launching in astrophysical sources, particularly transient systems, as discussed in Section~\ref{sec:obs_cons}.}

\subsection{General structure}
\label{sec:glob_stru}

Fig.~\ref{fig:field_lines} provides a more visual illustration of how magnetic field generation depends on disk thickness for $r\leq25r_g$. The figure shows poloidal slices of the accretion disk, with the $\varphi$-averaged toroidal magnetic field shown in color and the $\varphi$-averaged poloidal magnetic field lines shown as white streamlines. Panels (i, ii, iii) correspond to the thickest disk with $h/r\sim0.3$, panels (iv, v, vi) to the intermediate disk with $h/r=0.2$, and panels (vii, viii, ix) to the thinnest disk with $h/r=0.1$. Each simulation is shown at three different times: $t=10^3$ (i, iv, vii), $t=4\times10^3$ (ii, v, viii), and $t=3.7\times10^4$ (iii, vi, ix).

All simulations show increasing large-scale magnetic field organization as time progresses, although the degree of organization strongly depends on disk thickness.

The thickest disk reaches the most organized state of the 3 simulations. The $h/r\sim0.3$ simulation enters the MAD state and develops a clear large-scale vertical magnetic field threading the disk midplane\footnote{The large-scale poloidal field forms a closed magnetic loop, returning through the disk at larger radii, around $r\sim100\,r_g$, as required by the solenoidal condition.}. This strong poloidal field generates a toroidal magnetic field that changes sign above and below the midplane, due to shearing action on the large scale radial field. 

The intermediate-thickness disk ($h/r=0.2$) also develops a relatively organized magnetic structure, though less coherent than in the thickest case. In this simulation, the large-scale ordered field is primarily located above the disk, while the midplane remains dominated by disorganized turbulence. In this simulation, the black hole is threaded by a large-scale vertical magnetic field, consistent with the moderate value of $\phi_{\rm BH}$ shown in Fig.~\ref{fig:phi_BH} at $t=3.7\times10^4\,r_g/c$.

The thinnest disk develops the least organized magnetic field structure, with a  highly turbulent field both near and above the disk midplane. Nevertheless, a weak large-scale vertical magnetic field connected to the black hole is still present, consistent with its comparatively small value of $\phi_{\rm BH}$ at $t=3.7\times10^4\,r_g/c$.

\begin{figure}
\includegraphics[width=\linewidth]{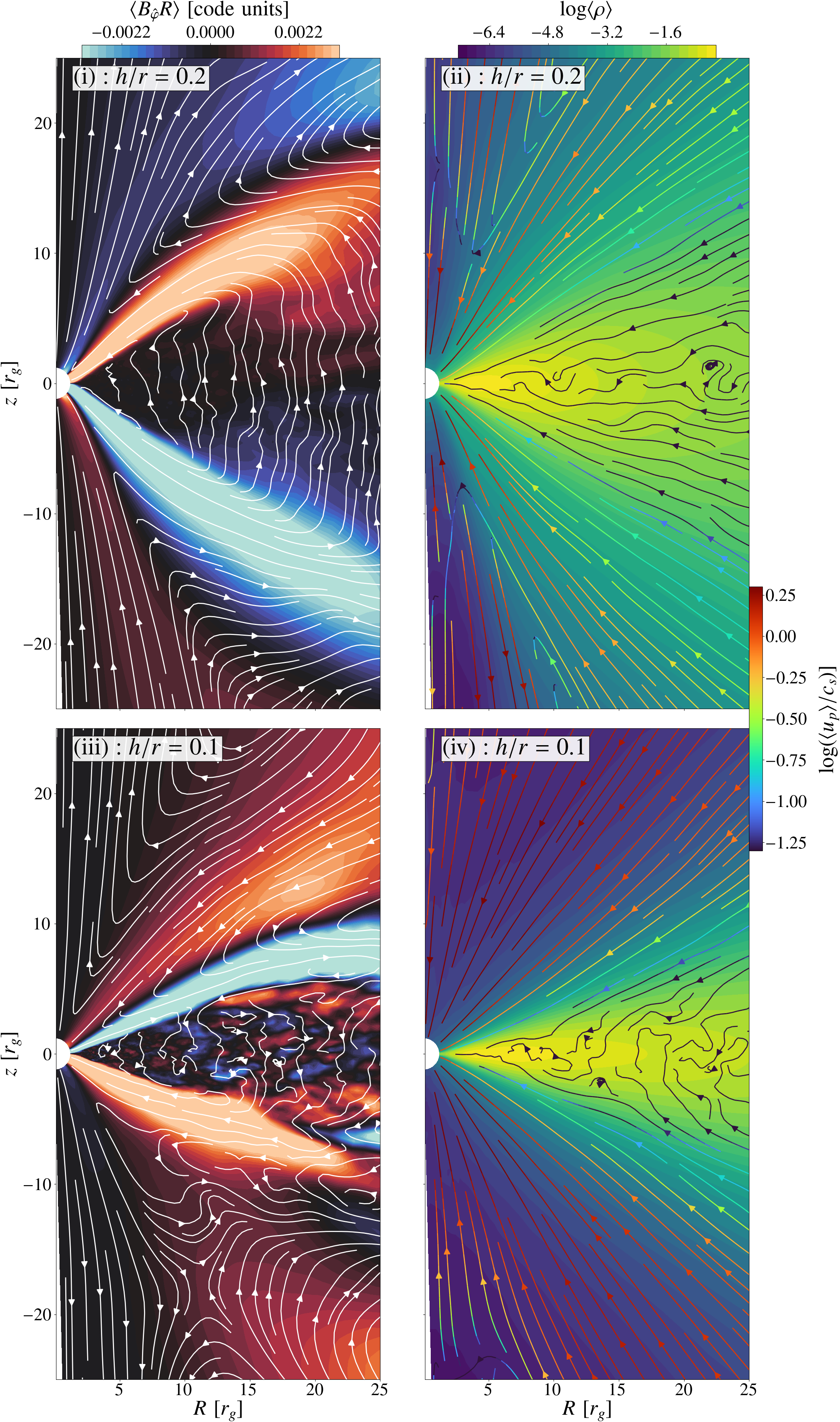}
    \caption{Time- and $\varphi$-averaged structure of the $h/r=0.2$ (i, ii) and $h/r=0.1$ (iii, iv) simulations. Panels (i, iii) show the toroidal magnetic field in colour with poloidal field lines as streamlines. Panels (ii, iv) show the logarithmic density with poloidal velocity streamlines coloured by their magnitude relative to the local sound speed. Both simulations develop a magnetically elevated configuration, with coherent large-scale magnetic fields and increasingly rapid inflow away from the dense, more slowly accreting midplane. This structure is weaker and less symmetric in the thinnest disk.}
    \label{fig:field_lines_avg}
\end{figure} 

{All simulations start with a purely toroidal magnetic field and pass through an evolutionary stage in which a poloidal field is present but not yet well organized, displaying relatively incoherent structure. These phases are accompanied by large-scale fluctuations in the toroidal magnetic field, which form regions of alternating polarity that propagate radially and vertically outward. In the thick-disk model, studied in detail by \citet{jacquemin-ide_magnetorotational_2024}, these structures remain coherent out to radii of $\sim50r_g$, with the toroidal field reversing polarity across these large scales. The intermediate-thickness simulation shows similar behavior, but with a shorter coherence length of $\sim10r_g$, while in the thinnest disk the radially propagating structures display a clearer wavelike character whose typical scale varies with radius. A more detailed discussion of these structures in the intermediate and thinnest disks is presented in our companion paper \citep{jacquemin-ide_dynamo_2026}}

While the simulations with $h/r\leq0.2$ do not reach the MAD state, they evolve toward a different configuration characterized by a weaker, yet still coherent, large-scale vertical magnetic flux. To understand the nature of this state, Fig.~\ref{fig:field_lines_avg}(i,iii) shows the time- and $\varphi$-averaged poloidal magnetic field lines, $\mean{B_p}$, together with the time- and $\varphi$-averaged toroidal field, $\mean{B_\varphi}$, for the $h/r=0.2$ (i) and $h/r=0.1$ (iii) simulations. Both simulations develop a large-scale vertical magnetic field threading the disk midplane, although its strength decreases strongly with disk thickness. As discussed in Section~\ref{sec:turbulent_behavior}, the corresponding mean poloidal plasma beta increases from $\beta_p\sim300$ for $h/r\sim0.3$ to $\sim7\times10^3$ for $h/r=0.2$ and $\sim4\times10^4$ for $h/r=0.1$. The two thinner disks also exhibit a strong, coherent toroidal field. For the $h/r=0.2$ disk, the toroidal field is mirror-symmetric about the midplane, whereas in the $h/r=0.1$ simulation this symmetry is preserved only within the disk ($|z|\lesssim5h$) and breaks down in the upper atmosphere. We attribute this loss of symmetry to the weaker large-scale field, which allows opposite-polarity loops to persist in the disk atmosphere.

The magnetic field configuration is strikingly similar to that of magnetically elevated disks reported in previous work \citep{zhu_global_2018,mishra_strongly_2020,jacquemin-ide_magnetic_2021}. Contrary to ours, those simulations were initialized with weak large-scale vertical magnetic fields ($\beta\gtrsim1000$) and evolve toward field geometries that closely resemble our $h/r=0.2$ simulation. The corresponding magnetic field strengths are also broadly consistent with the plasma betas measured in our thinner disks (see section \ref{sec:turbulent_behavior}). For the thinnest disk, this specific geometry is only visible within the northern hemisphere, in contrast to previous magnetically elevated disk simulations, which remain approximately symmetric about the midplane.

To further test this interpretation, Fig.~\ref{fig:field_lines_avg}(ii,iv) shows the poloidal velocity streamlines, colored by their magnitude normalized to the local sound speed, overlaid on the logarithm of the density. Both simulations display the characteristic signatures of magnetically elevated accretion: the radial inflow velocity increases with height, reaching transonic speeds for $h/r=0.2$ and supersonic speeds for $h/r=0.1$, at around a height of $z\sim6h$. However, the dense disk midplane remains comparatively turbulent and accretes more slowly, as expected for magnetically elevated disks \citep{jacquemin-ide_magnetic_2021}. 

\begin{figure}
	\includegraphics[width=\columnwidth]{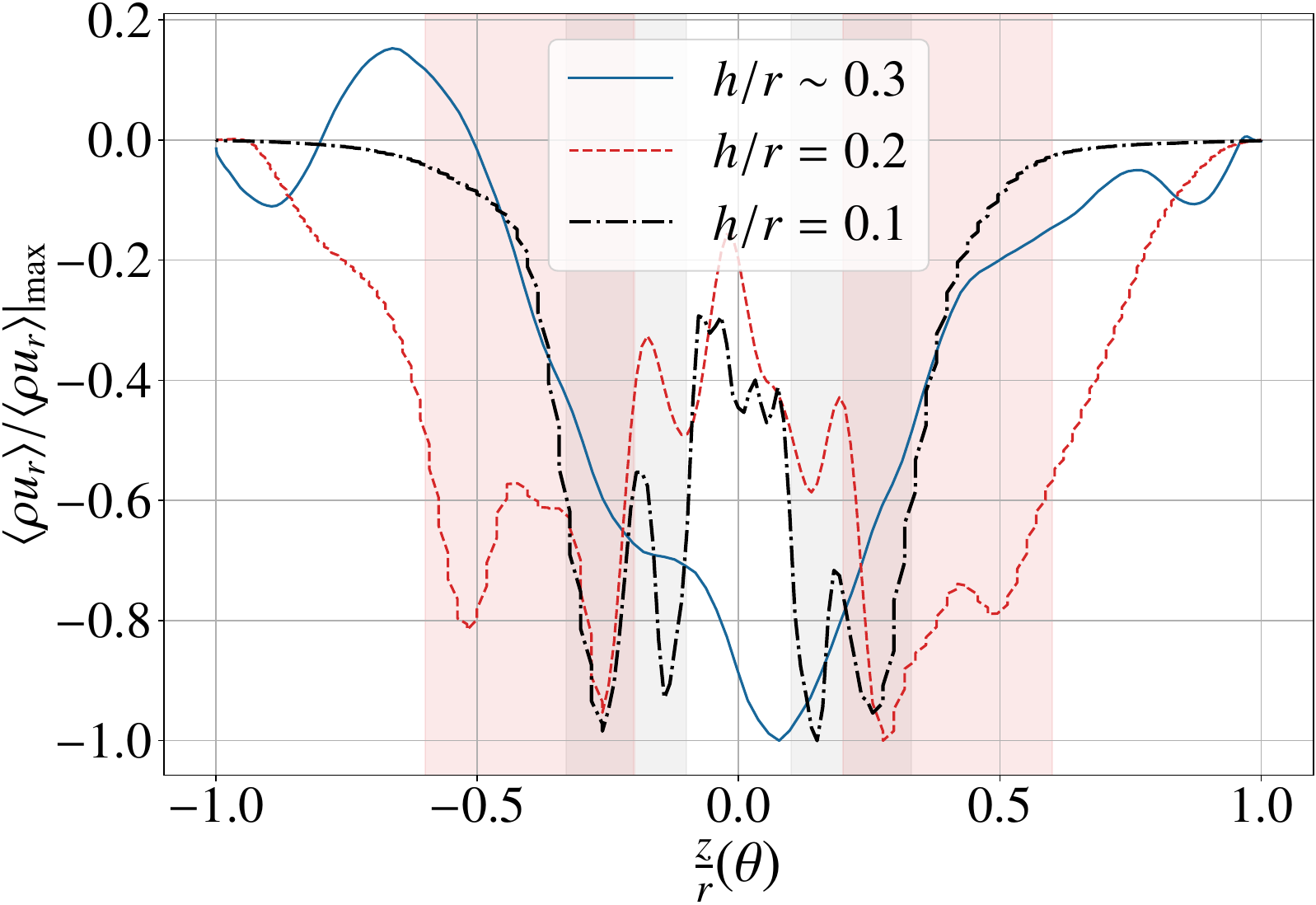}
    \caption{Normalized radial mass flux, $\mean{\rho u_r}$, as a function of height at $r=20r_g$, averaged between $t_1$ and $t_2$ for each simulation. The shaded regions mark $h\lesssim|z|\lesssim3h$ for the $h/r=0.2$ (red) and $h/r=0.1$ (grey) disks. In both thin disks, accretion is concentrated in layers above and below the midplane, where the radial mass flux exceeds its midplane value by more than a factor of three.}
    \label{fig:dens_flux}
\end{figure}

We also measure the radial mass flux, $\mean{\rho u_r}$, normalized to its maximum value, as a function of height at $r=20r_g$, as shown in Fig.~\ref{fig:dens_flux}. The quantities are averaged in time between $t_1$ and $t_2$; see Table~\ref{tab:simulations} for the corresponding time intervals. The red shaded region corresponds to $h\lesssim|z|\lesssim3h$ for the intermediate-thickness disk, while the grey shaded region marks the same region for the thinnest disk.

In both thin-disk simulations ($h/r\leq0.2$), the accretion flow is concentrated in two layers above and below the midplane, approximately at $h\lesssim|z|\lesssim3h$. The radial mass flux in these atmospheric layers is more than a factor of three larger than near the disk midplane. This contrast is particularly significant because the density decreases with height; the enhanced inflow velocities in these layers therefore compensate for the lower density, resulting in most of the mass accretion occurring away from the disk midplane. We find similar trends at all radii examined, $3r_g<r<30r_g$.

Simulations initialized with a weak large-scale vertical magnetic field also develop magnetically driven outflows \citep{zhu_global_2018,mishra_strongly_2020,jacquemin-ide_magnetic_2021}. We likewise find evidence of such outflows in the intermediate-thickness disk ($h/r=0.2$), although they are launched from significantly larger heights above the midplane, $|z|\geq20$ and $r\leq10$, than in \cite{jacquemin-ide_magnetic_2021}. This difference is likely due to our disk being roughly twice as thick. In contrast, we find no clear outflows in the thinnest disk ($h/r=0.1$). We attribute this to its much weaker and more disorganized large-scale poloidal magnetic field: the corresponding plasma beta is approximately an order of magnitude larger than in the weakest-field simulation of \cite{jacquemin-ide_magnetic_2021} (see Section~\ref{sec:turbulent_behavior}).

We therefore conclude that disks in which the large-scale poloidal magnetic field is generated self-consistently through dynamo action naturally evolve toward a magnetically elevated state \citep[see also][]{roberts_global_2025}. 

It is unclear if this magnetically elevated will continue to accumulate flux till it reaches the MAD state. Previous studies have shown that such disks eventually transition to the MAD state once they have had sufficient time to advect magnetic flux onto the black hole \citep{jacquemin-ide_magnetic_2021,wong_sane_2026}. Our simulations might be suggesting that the $h/r\leq0.2$ disks are following the same evolutionary pathway, but on a much longer timescale. We discuss this further in section \ref{sec:implications}.

\subsection{Local turbulent behavior}
\label{sec:turbulent_behavior}

To understand the growth and saturation stages of the different magnetic field components, we define different inverse plasma betas for every energetic component
\begin{align}
    \beta_\delta^{-1} &= \frac{\mean{\delta B^2}}{8 \pi \mean{P_0}},\\
    \beta_{\rm p}^{-1} &= \frac{\mean{ B_p}^2}{8 \pi \mean{P_0}},\\
    \beta_{\rm \theta}^{-1} &= \frac{\mean{ B_\theta}^2}{8 \pi \mean{P_0}},\\
    \beta_\varphi^{-1} &= \frac{\mean{ B_\varphi}^2}{8 \pi \mean{P_0}},
\end{align}
and the turbulent angular momentum transport coefficient
\begin{equation}
    \alpha_{\nu} = \frac{\mean{\rho \delta u_r \delta u_\varphi}-\mean{\delta B_\varphi \delta B_r}}{4 \pi \mean{P_0}},
\end{equation} 
where $P_0$ is the pressure measure at the disk midplane. Note that that the $\mean{B_p}^2$ captures the large-scale poloidal field generated by the MRI dynamo rather than the small-scale turbulent poloidal flow. In the same way $\mean{B_\varphi}$ also captures the large scale toroidal field which is a driver but also a consequence of the MRI dynamo. 

\begin{figure}
	\includegraphics[width=\columnwidth]{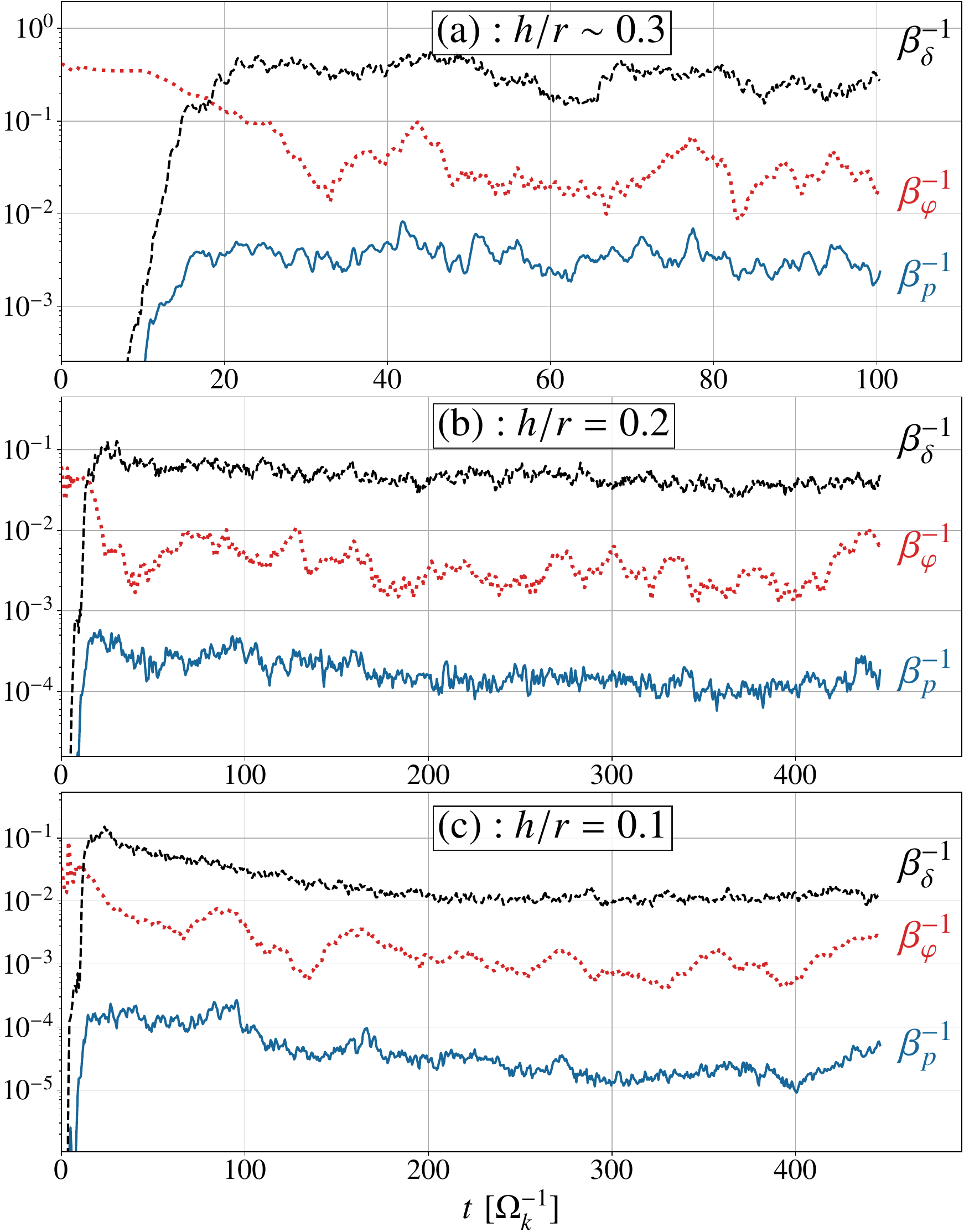}
    \caption{Temporal evolution of the inverse plasma betas associated with the turbulent, mean toroidal, and mean poloidal magnetic fields at $r=20r_g$, shown in units of the shear timescale $\Omega_k^{-1}$ for $h/r\sim0.3$ (a), $0.2$ (b), and $0.1$ (c). All simulations show rapid exponential growth of the turbulent and mean poloidal fields, followed by relaxation toward a statistical quasi-steady state. The saturated fields satisfy $\mean{\delta B^2}\gg\mean{B_\varphi}^2\gg\mean{B_p}^2$, with the separation between the mean toroidal and poloidal components increasing toward thinner disks.}
    \label{fig:betas_time}
\end{figure}

We show in Fig.~\ref{fig:betas_time} the temporal evolution of the inverse plasma betas associated with the different magnetic field components, as a function of time in units of the shear timescale\footnote{Not to be confused with orbital periods, which differ by a factor of $2\pi$.} $\Omega_k^{-1}$. The quantities are vertically averaged within the disk, see Eq.~\ref{Eq:disk_avg} (with averages computed before division) and evaluated\footnote{We find very similar trends at all radii examined, $5r_g\leq r\leq35r_g$. In particular, the relative separation between the different magnetic-field components remains similar, while their absolute strengths increase toward larger radii, where the relaxation phase described below has not yet been completed.} at $r=20r_g$ for the three disk thicknesses: $\mathrm{(a):}\,h/r\sim0.3$, $\mathrm{(b):}\,h/r=0.2$, and $\mathrm{(c):}\,h/r=0.1$. The reason why $\beta_\varphi^{-1}(t=0)$ is slightly smaller than the initial value $\beta_{\rm ini}^{-1}=1/5$ is that the inverse plasma beta decreases sharply beyond a few scale heights;  averaging over the vertical extent of the disk therefore lowers its effective value. We caution the reader that the temporal axes differ between panels because the thickest disk evolves more rapidly, with the loop that leads to the transition to the MAD state connecting with the BH at $t=10^{4}\simeq100\,\Omega_k^{-1}(R=20\,r_g)$.

In all simulations, we observe an exponential growth of both the turbulent magnetic field, $\delta\beta^{-1}$, and the mean large-scale poloidal field, $\beta_p^{-1}$. This growth occurs rapidly, within $\sim10\Omega_k^{-1}$, corresponding to roughly ten linear MRI growth times. The exponential amplification of $\delta\beta^{-1}$ and $\beta_p^{-1}$ is followed by a slower relaxation phase, during which the fields decrease by a factor of a few in the intermediate-thickness case (Fig.~\ref{fig:betas_time}b) and by nearly an order of magnitude in the thinnest disk case (Fig.~\ref{fig:betas_time}c). In the thickest disk, the MAD state is reached so quickly that this relaxation phase is barely visible in Fig.~\ref{fig:betas_time}(a). 
The $\varphi$-averaged toroidal magnetic field, $\beta_\varphi^{-1}$, decreases once the turbulent field becomes dynamically important, i.e. when $\mean{\delta B^2}\sim\mean{B_\varphi}^2$. This behavior can be understood as the turbulent feedback acting on the mean magnetic field.

Once the magnetic fields reach a statistical quasi-steady state, we observe a clear energetic ordering between the different components:
\begin{equation}
    \mean{\delta B^2}\gg\mean{B_\varphi}^2\gg\mean{B_p}^2.
\end{equation} 
For all simulations, we consistently find $\delta\beta^{-1}/\beta_\varphi^{-1}\sim10$. However, the ratio $\beta_\varphi^{-1}/\beta_p^{-1}$ increases as the disk becomes thinner, evolving from $\sim10$ in the thickest disk ($h/r\sim0.3$, Fig.~\ref{fig:betas_time}a), to $\sim10^{1.5}$ for the intermediate case ($h/r=0.2$, Fig.~\ref{fig:betas_time}b), and finally to $\sim10^2$ for the thinnest disk ($h/r=0.1$, Fig.~\ref{fig:betas_time}c).
 
\begin{figure}
	\includegraphics[width=\columnwidth]{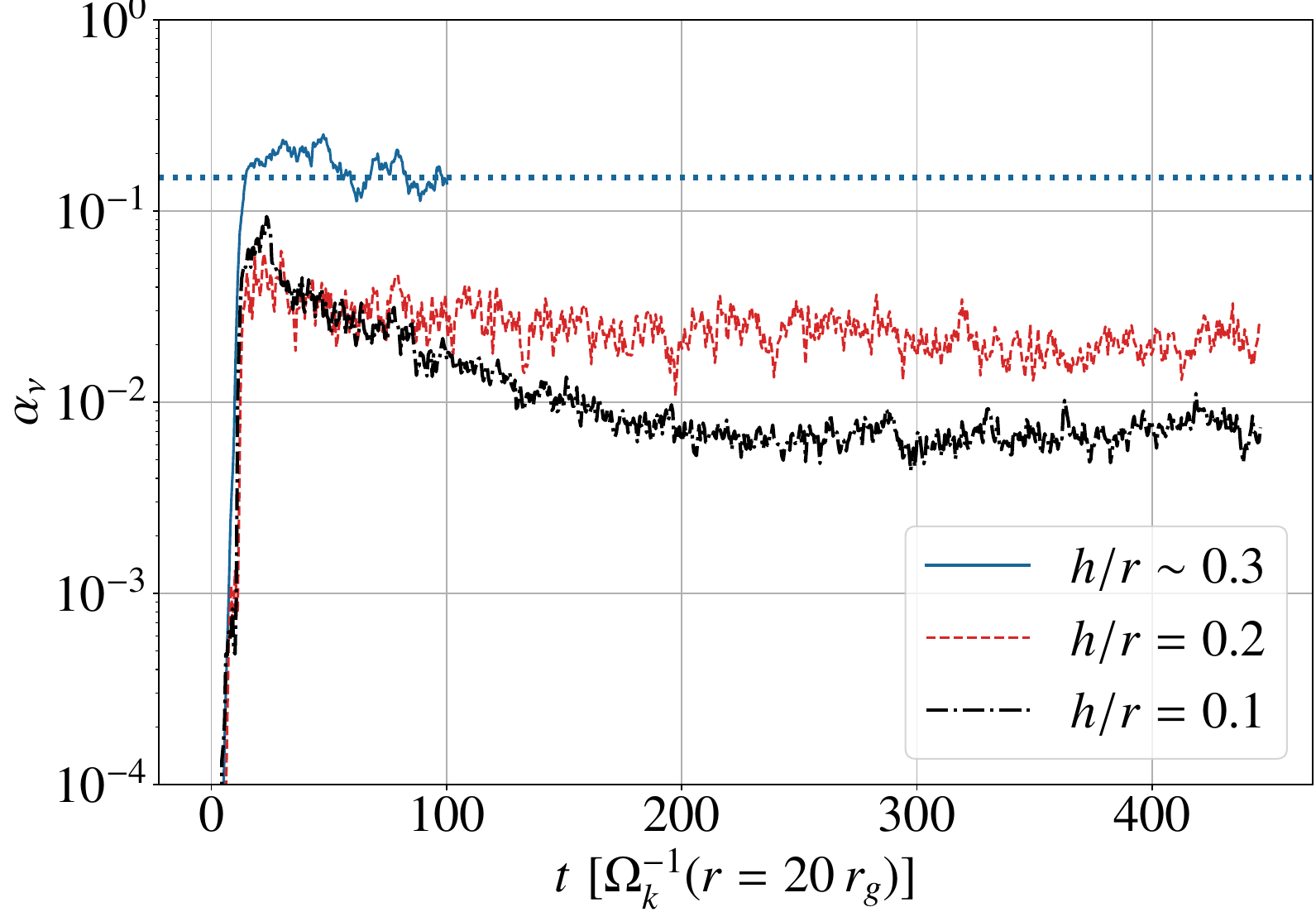}
    \caption{Temporal evolution of the turbulent stress coefficient, $\alpha_\nu$, in units of the shear timescale $\Omega_k^{-1}$ for the three disk thicknesses. The turbulence grows on a similar timescale in all simulations, but its saturated strength decreases strongly with disk thickness, from $\alpha_\nu\simeq1.3\times10^{-1}$ for $h/r\sim0.3$ to $8\times10^{-3}$ for $h/r=0.1$.}
    \label{fig:alpha}
\end{figure} 

To better compare the strength of the magnetic turbulence between simulations, we show in Fig.~\ref{fig:alpha} the evolution of $\alpha_\nu$ as a function of time in units of the shear timescale $\Omega_k^{-1}$ for the three disk thicknesses. The figure shows that the turbulence grows on a similar timescale in all simulations. The turbulent stress coefficient $\alpha_\nu$ follows an evolution similar to that of the turbulent magnetic energy, $\delta\beta^{-1}$. 

We find a clear dependence of the turbulence strength on disk thickness: thinner disks develop weaker turbulence, with $\alpha_\nu$ decreasing from $1.3\times10^{-1}$ for $h/r\sim0.3$, to $2\times10^{-2}$ for $h/r=0.2$, and finally to $8\times10^{-3}$ for the thinnest disk with $h/r=0.1$. 

\begin{figure}
	\includegraphics[width=0.99\columnwidth]{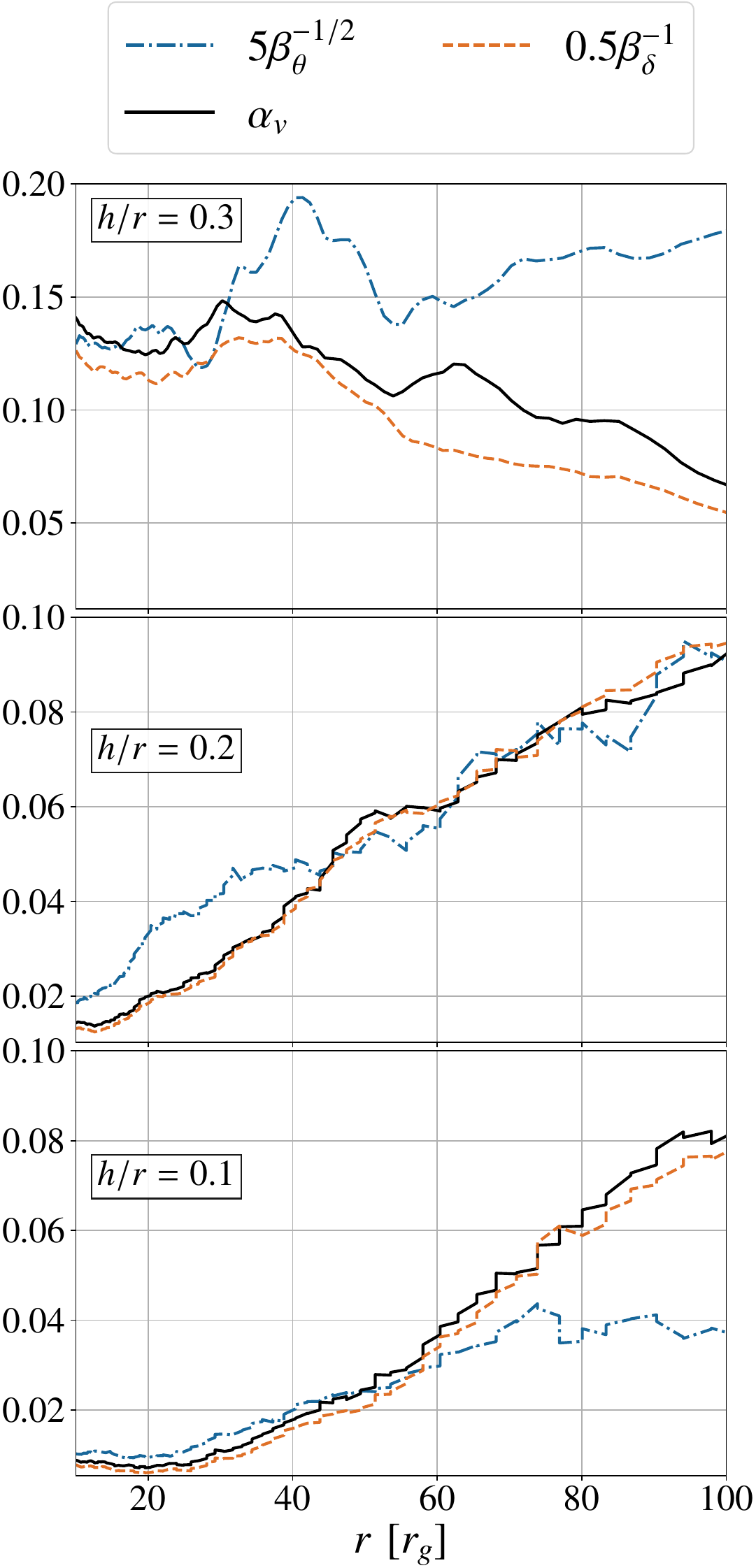}
    \caption{Radial profiles of the turbulent stress parameter, $\alpha_\nu$, together with the scalings $0.5\beta_\delta^{-1}$ and $5\beta_\theta^{-1/2}$ for the three disk thicknesses. The turbulent-field scaling closely follows the measured stress in all simulations, while the vertical-field scaling also provides a good description, particularly in the thinner disks. This supports the interpretation that the MRI turbulence is primarily regulated by the large-scale vertical magnetic field.}
    \label{fig:test_law}
\end{figure}

In Fig.~\ref{fig:test_law}, we show the turbulent stress parameter $\alpha_\nu$ as a function of radius, averaged over the disk and in time between $t_1$ and $t_2$ (see Table \ref{tab:simulations}). For the thickest-disk simulation, we find a relatively flat profile out to $r=60r_g$, which is likely related to the lack of convergence at larger radii. In the two thinner-disk simulations, $\alpha_\nu$ instead increases with radius, with a relatively flat profile only at $r\lesssim20r_g$. This behavior is related to the slower relaxation of these simulations, during which we observe a decrease in the turbulent field strength (see Fig.~\ref{fig:alpha} and Fig.~\ref{fig:betas_time}). In particular, the outer regions have not had sufficient time to relax to their final state, which is characterized by a lower $\alpha_\nu$.

MRI turbulence is expected to satisfy the following scaling between the turbulent stress parameter and the turbulent magnetic field strength:
\begin{equation}
\alpha_\nu \simeq 0.5\beta_\delta^{-1}.
\end{equation}
We show this scaling in Fig.~\ref{fig:test_law}, evaluated as a function of radius, with the quantities averaged over the disk and in time between $t_1$ and $t_2$ shown in Table \ref{tab:simulations}. We find very good agreement with the numerical results; indeed, as expected for MRI turbulence, $\alpha_\nu\beta_\delta\sim0.5$ \citep{hawley_local_1995,lesur_magnetohydrodynamics_2021}.

Local shearing-box simulations have established scalings relating the strength of the mean vertical magnetic field, characterized by $\beta_\theta$, to the turbulent stress $\alpha_\nu$. To test whether the vertical magnetic field regulates the turbulent stress in our simulations, we compare $\alpha_\nu$ with
\begin{align}
\label{eq:vert_MRI_scale}
\alpha_\nu^{B_\theta} &\simeq C_{B_\theta}\beta_\theta^{-1/2}.
\end{align}
We adopt $C_{B_\theta}=5$, which provides the best agreement with our simulations and is also approximately intermediate between results from stratified and non-stratified shearing-box simulations \citep{hawley_local_1996,salvesen_accretion_2016,lesur_magnetohydrodynamics_2021}\footnote{Stratified shearing-box simulations find $C_{B_\theta}\simeq10$, while non-stratified simulations find $C_{B_\theta}\simeq3$. The scaling in stratified simulations is obtained by averaging over the entire numerical domain, whereas here we focus on the region near the disk midplane. It is therefore unclear how directly these averages can be compared.}

We show this scaling as a function of radius in Fig.~\ref{fig:test_law}. The quantities are integrated over the disk and averaged in time between $t_1$ and $t_2$, with the corresponding intervals listed in Table~\ref{tab:simulations}. Overall, $5\beta_\theta^{-1/2}$ approximately reproduces the measured turbulent stress, particularly at smaller radii and in the thinner disks.

For the thinnest disk ($h/r=0.1$), the scaling agrees with $\alpha_\nu$ to within $\sim20\%$ out to $r=60r_g$. For the intermediate-thickness disk, it provides a reasonable description at smaller radii, with discrepancies of up to $\sim30\%$ for $r\lesssim30r_g$, and agrees even more closely with the measured stress at larger radii.

The thickest disk shows poorer agreement with the vertical-field MRI scaling, with discrepancies of $\sim20$--$60\%$ for $r\leq60r_g$. Taken together, these comparisons suggest that the strength of the MRI turbulence is primarily regulated by the large-scale vertical magnetic field. The correspondence between $\alpha_\nu$ and the vertical-field scaling therefore supports a picture in which the dynamo-generated poloidal field sets the overall turbulent state of the disk.

{In this section, we have identified two dependencies on $h/r$ that directly affect the rate at which large-scale magnetic flux accumulates onto the black hole (see Fig.~\ref{fig:phi_BH}). First, the overall strength of the turbulence decreases as $h/r$ decreases. The agreement with the vertical-MRI scaling suggests that this weakening is closely linked to the decrease in the large-scale vertical magnetic field, with the MRI turbulence appearing to be regulated by this field. Second, the separation between the turbulent and mean magnetic fields increases with decreasing $h/r$, with $\mean{\delta B^2}/\mean{B_p}^2$ growing from $\sim10^2$ for $h/r\sim0.3$, to $\sim5\times10^2$ for $h/r=0.2$, and finally to $\sim10^3$ for $h/r=0.1$. This trend follows the scaling of $\beta_\varphi^{-1}/\beta_p^{-1}$, since the ratio involving the turbulent and toroidal fields remains approximately constant with $h/r$. Taken together, these results suggest that the strength of the MRI turbulence is closely tied to the large-scale vertical field generated by the dynamo, with weaker mean fields in thinner disks producing correspondingly weaker turbulence. We examine how this large-scale field is generated in Section~\ref{sec:field_gen}.}

\section{Field generation}
\label{sec:field_gen}

In this section, we investigate how the large-scale axisymmetric magnetic field is generated and sustained as a function of disk thickness, considering first the toroidal and then the poloidal field. Local zero-net-flux simulations have shown that nonlinear interactions involving non-axisymmetric MRI fluctuations can sustain a large-scale axisymmetric poloidal field, which is subsequently sheared into a toroidal field by differential rotation \citep{lesur_self-sustained_2008,herault_periodic_2011,riols_global_2013,riols_dissipative_2015,mamatsashvili_zero_2020}. Our simulations introduce an additional complexity, as they self-consistently generate regions of coherent vertical magnetic flux from an initially toroidal field.

The emergence of this vertical flux feeds back on both the small-scale turbulence and the large-scale field evolution. As shown in the previous section, the vertical field modifies the MRI turbulence and its associated transport, potentially altering the turbulent electromotive forces responsible for generating the large-scale field. In addition, it permits axisymmetric MRI motions, including channel-like structures, that can contribute directly to the evolution of the axisymmetric poloidal field \citep{goodman_parasitic_1994,latter_mri_2010}. Although the emergence of vertical flux allows axisymmetric MRI motions to contribute directly to the poloidal-field evolution, studies of local net-vertical-flux MRI turbulence indicate that non-axisymmetric fluctuations can dominate the turbulent energy and angular momentum transport, even while channel modes remain dynamically important \citep{longaretti_mri-driven_2010,gogichaishvili_active_2018}. The relative importance of these motions for large-scale poloidal-field generation, however, remains unclear. Moreover, large-scale magnetic field reversals persist in stratified simulations with weak net vertical flux \citep{salvesen_accretion_2016}. We therefore investigate both the direct contribution of axisymmetric motions and the nonlinear transfer from non-axisymmetric fluctuations, without attempting to separate the two mechanisms completely.

\subsection{Toroidal magnetic field budget}
\begin{figure}
	\includegraphics[width=0.9\columnwidth]{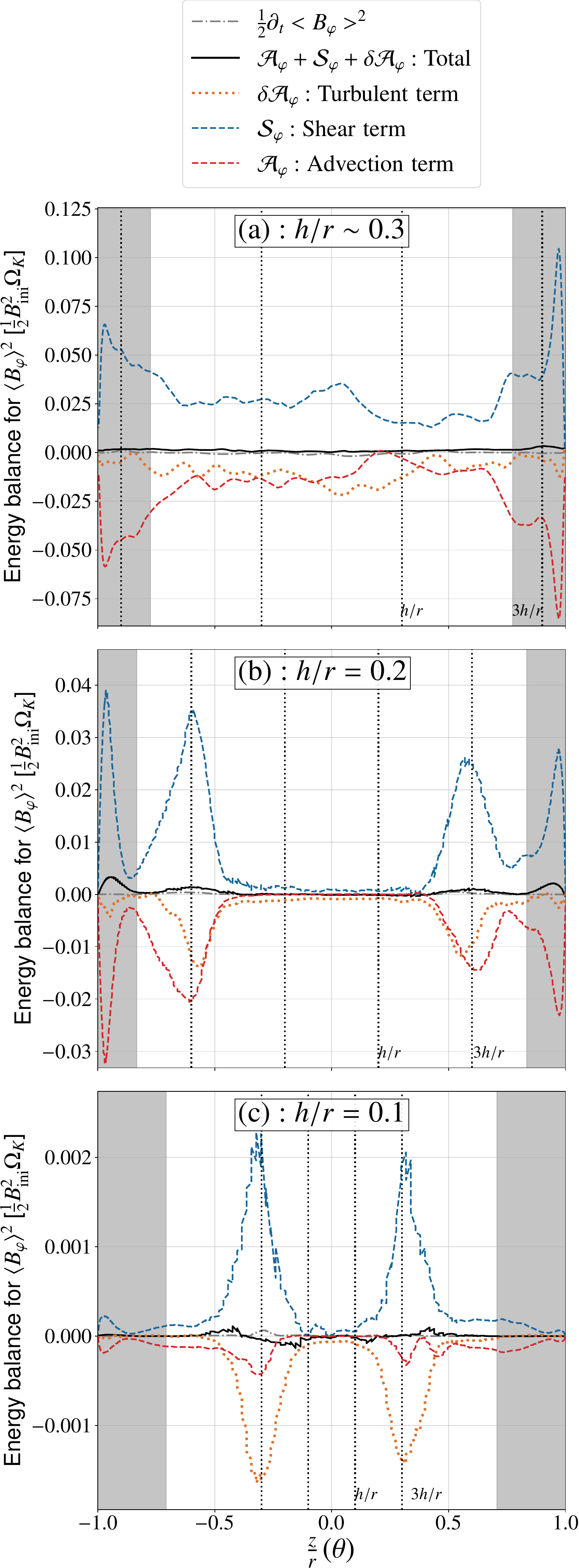}
    \caption{Vertical profiles of the toroidal magnetic-energy budget at $r=20r_g$ for $h/r\sim0.3$ (a), $0.2$ (b), and $0.1$ (c), showing the shear, mean-advection, and turbulent terms from Eq.~\ref{Eq:tor_ener}, together with their sum and $\frac{1}{2}\partial_t\mean{B_\varphi^2}$. The close agreement between the latter two quantities indicates a local quasi-steady balance. Shear generates toroidal magnetic energy, while turbulent and axisymmetric terms remove it. In the thinner disks, both generation and dissipation are suppressed near the midplane and become concentrated around $|z|\sim3h$.}
    \label{fig:toroidal_gen}
\end{figure} 

Following \cite{jacquemin-ide_magnetorotational_2024} we derive an equation for the generation of the toroidal magnetic field, by taking the Reynolds averaged induction equation (Eq.~\ref{Eq:avg_induct}) and dotting it with respect the toroidal magnetic field. This yields
\begin{equation}
    \frac{1}{2}\pdv{\mean{B_\varphi}^2}{t} = \mathcal{S}_\varphi+\mathcal{A}_\varphi+\delta\mathcal{A}_\varphi,
    \label{Eq:tor_ener}
\end{equation}
where
\begin{equation}
    \mathcal{S}_\varphi = \vmean{B_\varphi}\cdot\rot\left(\vmean{u_\varphi}\times\vmean{B_p}\right)
\end{equation}
is the effect of the large scale shear, which is in part the driven by the shear of the poloidal field ($\Omega$-effect) but could also be tied to axisymmetric MRI channel mode action,
\begin{equation}
    \mathcal{A}_\varphi = \vmean{B_\varphi}\cdot\rot\left(\vmean{u_p}\times\vmean{B_\varphi}\right) 
\end{equation}
is the axisymmetric induction or transport  of the large scale toroidal field by the axisymmetric poloidal fluid flow, and
\begin{equation}
    \delta \mathcal{A}_\varphi = \vmean{B_\varphi}\cdot\rot\left(\mathbf{\mathcal{E}_p}\right)
\end{equation}
is the turbulent generation or dissipation of the large scale magnetic field by the turbulent motions, the coherent feedback on the mean toroidal field.

We show in Fig.~\ref{fig:toroidal_gen}(a--c) the different terms in Eq.~\ref{Eq:tor_ener} as functions of $z/r(\theta)$, normalized to the initial toroidal magnetic energy and the local Keplerian shear rate, for the three disk thicknesses: (a) $h/r\sim0.3$, (b) $h/r=0.2$, and (c) $h/r=0.1$. All terms are time averaged between $t_1$ and $t_2$ (see Table \ref{tab:simulations}) and evaluated at $r=20r_g$. We have verified that similar trends are found at other radii, $5r_g\leq r\leq35r_g$. We also show the time derivative of the toroidal magnetic energy, $\frac{1}{2}\partial_t\mean{B_\varphi^2}$, together with the sum of the three magnetic-field generation terms, i.e., all terms on the right-hand side of Eq.~\ref{Eq:tor_ener}. For all three simulations, these two quantities are nearly identical and close to zero. This indicates that we have accounted for the relevant sources and sinks of toroidal magnetic energy and that the toroidal field is in a local quasi-steady state, with magnetic energy generation balanced by dissipation. This quasi-steady state does not, however, imply that the system is globally stationary, as the large-scale advection of poloidal magnetic flux continues to drive the global evolution; see Section~\ref{sec:field_transport}.

For all simulations, we restrict our analysis to $|\theta-\pi/2|\leq\pi/4$, corresponding to $|z/r|\leq0.7$. At larger heights, the signal is dominated by the magnetic funnel associated with field lines connected to the black hole. The magnetic field evolution in this region is primarily an expression of Ferraro's law of isorotation and is therefore not relevant to the disk dynamo considered here. We mark these regions in grey in Fig.~\ref{fig:toroidal_gen}.

Within this region, we find similar qualitative trends across all three simulations. Magnetic energy is generated by the shear term, $\mathcal{S}_\varphi$, while the turbulent term, $\delta\mathcal{A}_\varphi$, dissipates magnetic energy and the axisymmetric induction term, $\mathcal{A}_\varphi$, also removes toroidal magnetic energy. The main difference between the intermediate and thinnest disks is the relative importance of these two loss terms: $\mathcal{A}_\varphi$ dominates over $\delta\mathcal{A}_\varphi$ in the intermediate case, while the opposite is true in the thinnest disk.

{We first consider the shear contribution, which can be further decomposed as}
{\scriptsize
\begin{equation}
\begin{aligned}
    \mathcal{S}_\varphi
    ={}&
    \underbrace{
    \mean{B_\varphi}\mean{B_r}
    r\partial_r{\frac{\mean{u_\varphi}}{r}}
    }_{\text{radial shear}}
    +
    \underbrace{
    \frac{\mean{B_\varphi}\mean{B_\theta}}{r}\sin\theta
    \partial_\theta\frac{\mean{u_\varphi}}{\sin\theta}
    }_{\text{latitudinal shear}} .
\end{aligned}
\label{eq:shear_phi}
\end{equation}
}
{The generation of the toroidal magnetic field is dominated in all three simulations by the radial shear associated with differential rotation. In the thinnest and thickest disks, the latitudinal shear contribution is smaller by roughly a factor of \(20\), while in the intermediate-thickness disk it reaches about one-sixth of the radial contribution. This suggests that both ordinary winding of the radial field and axisymmetric MRI motions may contribute, but the weakness of the latitudinal shear term indicates that channel-like motions are unlikely to dominate the toroidal-field generation. Overall, \(B_\varphi\) is produced primarily by the radial \(\Omega\)-effect, while latitudinally varying axisymmetric motions play a more important role in the evolution of the poloidal field (see next section).}

{We next turn to $\mathcal{A}_\varphi$, which describes the effect of the mean poloidal flow on the toroidal magnetic field,}

{\scriptsize
\begin{equation}
\begin{aligned}
    \mathcal{A}_\varphi
    ={}&
    \underbrace{-\mean{B_\phi}\mean{u_r}\,\partial_r\mean{B_\phi}}_{\text{radial advection}}
    \underbrace{-\frac{\mean{B_\phi}\mean{u_\theta}}{r}\,\partial_\theta\mean{B_\phi}}_{\text{latitudinal advection}}
    \\
    &\underbrace{-\mean{B_\phi}^{\,2}\,\nabla\cdot\mean{\mathbf{u}_p}}_{\text{compression}}
    +\underbrace{\mean{B_\phi}^{\,2}
    \left(
        \frac{\mean{u_r}}{r}
        +
        \frac{\mean{u_\theta}\cot\theta}{r}
    \right)}_{\text{geometric}} .
\end{aligned}
\label{eq:advec_phi}
\end{equation}
}
{In all three simulations, the contribution of the mean poloidal flow is dominated by radial advection, which acts to reduce the local toroidal magnetic energy. The other contributions do not compensate for this loss, leaving differential rotation as the dominant source of $B_\varphi$.}

The main difference between the simulations lies in the vertical profiles of the field-generating and dissipating terms. For ease of comparison, Fig.~\ref{fig:toroidal_gen} marks the heights $z=h$ and $z=3h$. In the thickest-disk simulation, all quantities exhibit relatively flat profiles, with axisymmetric/large-scale magnetic field generation and dissipation having comparable magnitudes throughout the disk and up to the transition into the funnel. In contrast, for the thinner disks ($h/r\leq0.2$), both axisymmetric/large-scale magnetic field generation and dissipation are strongly suppressed at heights $|z|\lesssim2h$ and reach their maximum around $|z|\sim3h$. Thus, in the thinner disks, axisymmetric/large-scale magnetic-field generation  and dissipation are concentrated in relatively narrow layers above and below the disk midplane, rather than being distributed throughout the disk body. This may contribute to the decrease in both the turbulent stress, $\alpha_\nu$, and the strength of the mean magnetic field with decreasing disk thickness. As the disk becomes thinner, magnetic field generation is  confined to a smaller fraction of the disk volume. {This stratification may also be related to the emergence of the magnetically elevated disk structure in the two thinner simulations ($h/r\leq0.2$), suggesting that the vertically localized field-generation processes identified here may contribute to establishing that configuration.}

The rate of large-scale field generation is also affected by the disk thickness. 
Once normalized to the initial toroidal magnetic energy, $B_{\rm ini}$, we can notice a decline in all the terms with decreasing disk thickness. Indeed, the magnitude of the field generation rate decreases from $0.03$ for $h/r\sim0.3$, to $0.002$ for $h/r=0.1$. However, the intermediate disk thickness simulation, $h/r=0.2$, shows a similar average magnitude for the field generation rate, i.e. $0.03$,  albeit as described above this is over a smaller solid angle, as field generation is confined vertically for the intermediate disk thickness case. 

\subsection{Poloidal magnetic field budget}

We have shown that sustaining the toroidal field requires the shear of a large-scale/axisymmetric poloidal field. Therefore, the generation of the poloidal field is essential to the large-scale field generaion mechanism.

In a similar manner, we derive an evolution equation for the mean poloidal magnetic energy,
\begin{equation}
    \frac{1}{2}\pdv{\mean{B_p}^2}{t} = \mathcal{A}_p+\delta\mathcal{A}_p,
    \label{Eq:pol_energ}
\end{equation}
where
\begin{equation}
    \mathcal{A}_p =\mathcal{A}_r + \mathcal{A}_\theta =  \vmean{B_p}\cdot\rot\left(\vmean{u_p}\times\vmean{B_p}\right)
    \label{eq:advec_pol}
\end{equation}
represents the advection and induction of the large-scale poloidal magnetic field by the axisymmetric poloidal flow, and
\begin{equation}
    \delta \mathcal{A}_p = \vmean{B_p}\cdot\rot\left(\mathbf{\mathcal{E}_\varphi}\mathbf{e_\varphi}\right)
    \label{eq:delAp_first}
\end{equation}
represents the turbulent generation and dissipation of the mean poloidal magnetic field.

We first focus on the turbulent term, $\delta\mathcal{A}_p$. As shown by \cite{jacquemin-ide_magnetorotational_2024}, this term contains both positive and negative contributions to the mean poloidal magnetic energy. To disentangle these competing effects, we decompose $\delta\mathcal{A}_p$ into contributions from individual azimuthal Fourier modes. 
First we decompose the poloidal velocity and magnetic fields into azimuthal Fourier modes,
\begin{align}
\mathbf{u_p} &= \sum_{m=-\infty}^{\infty}\mathbf{u_p}^{m}e^{im\varphi},\\
\mathbf{B_p} &= \sum_{m=-\infty}^{\infty}\mathbf{B_p}^{m}e^{im\varphi},
\label{Eq:BpFour}
\end{align}
where $m$ is the azimuthal wavenumber and 
the spectral coefficients $\mathbf{u_p}^{m}$ and $\mathbf{B_p}^{m}$ depend only on $r$ and $\theta$.

\begin{figure}
	\includegraphics[width=\columnwidth]{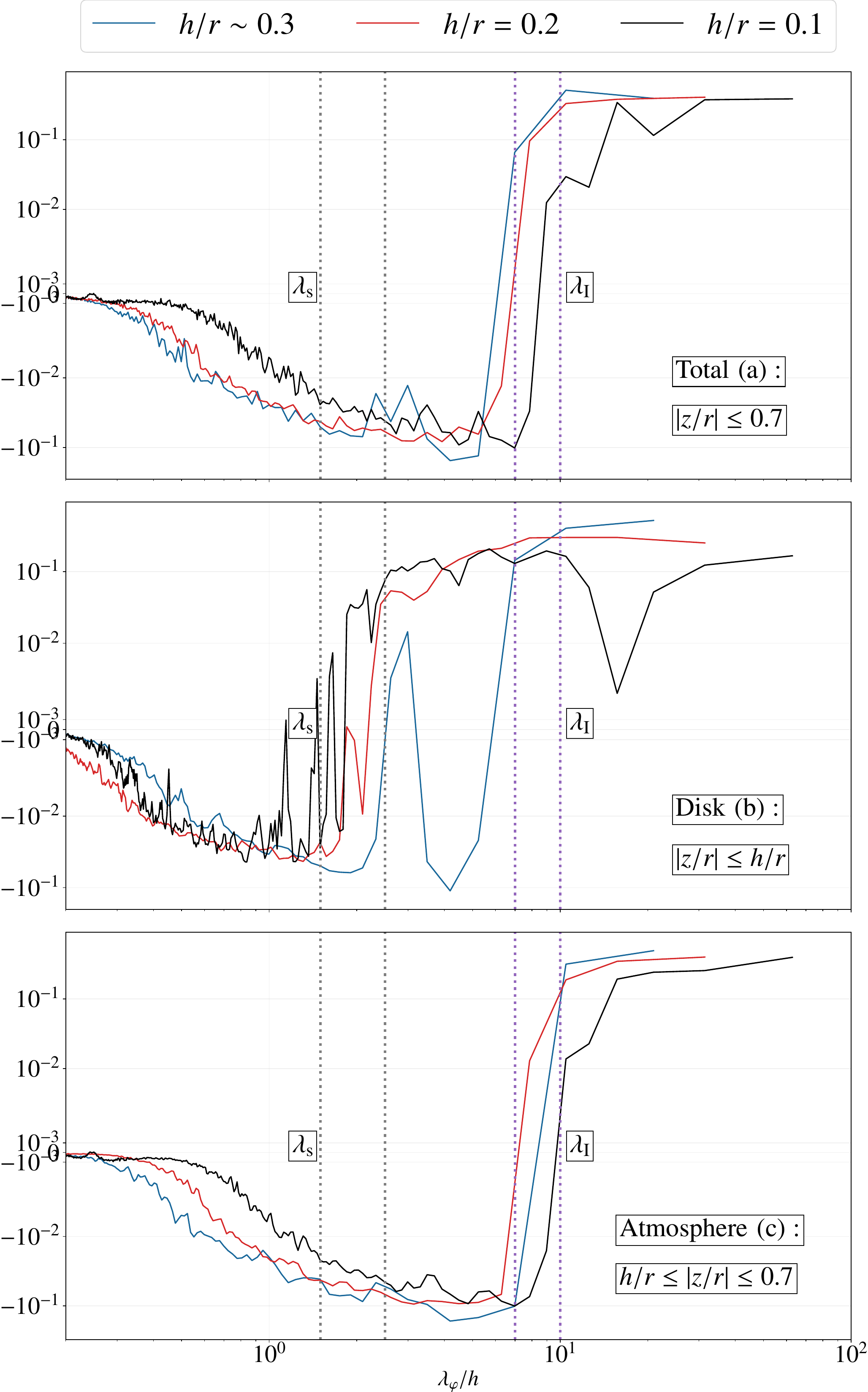}
    \caption{Normalized spectrum of the poloidal-field generation term, $\delta\mathcal{A}_p^{(m)}$, as a function of azimuthal wavelength $\lambda_\varphi/h$ at $r=20r_g$. Panels show averages over the full vertical region (a), the disk midplane (b), and the disk atmosphere (c). Large-scale modes with $\lambda_\varphi>\lambda_I$ generate poloidal magnetic field in both regions, intermediate modes with $\lambda_S<\lambda_\varphi<\lambda_I$ generate field within the disk but dissipate it in the atmosphere, and small-scale modes with $\lambda_\varphi<\lambda_S$ dissipate the field in both regions. The characteristic transition wavelengths $\lambda_I$ and $\lambda_S$ are shown as vertical dashed lines and are similar across disk thicknesses when normalized by $h$, indicating that the dynamo scale hierarchy is set primarily by the local disk thickness.}
    \label{fig:poloidal gen}
\end{figure} 

Applying Parseval's theorem, the turbulent EMF can be decomposed into contributions from individual azimuthal modes,
\begin{equation}
    \mathcal{E}_\varphi = \mean{\delta \mathbf{u_p}\times\delta \mathbf{B_p}}\cdot\mathbf{e_\varphi} = \sum\limits_{m=1}^\infty\mathcal{E}_\varphi^{(m)}
\end{equation}
where $\mathbf{e_\varphi}$ is the unit vector along the azimuthal direction and 
\begin{equation}
   \mathcal{E}_\varphi^{(m)} = 2\mathcal{R}\left[\mathbf{u_p}^{m}\times \left(\mathbf{B_p}^{m}\right)^{*}\right]\cdot\mathbf{e_\varphi}
    \label{eq:EMF_mode}
\end{equation}
{is the axisymmetric EMF produced by the nonlinear interaction between the $m$ and $-m$ Fourier components, with $\mathcal{R}$ denoting the real part and $^*$ the complex conjugate.\footnote{We do not Fourier transform the EMF directly. Instead, we Fourier transform the velocity and magnetic fields and reconstruct $\mathcal{E}_\varphi^{(m)}$ using Eq.~(\ref{eq:EMF_mode}).} We use the superscript $(m)$ to denote the contribution of the $\pm m$ mode pair to the axisymmetric EMF, distinguishing it from the superscript $m$, which denotes an individual Fourier coefficient. The sum begins at $m=1$ because the $m=0$ contribution corresponds to the axisymmetric fields\footnote{Indeed, $\mean{X_p}=X_p^{m=0}$.} and is already included in the advection term $\mathcal{A}_p$ in Eq.~(\ref{eq:delAp_first}).}

The feedback of the EMF on the energetics of the poloidal field can then be written as
\begin{equation}
    \delta \mathcal{A}_p = \sum\limits_{m=1}^\infty\delta\mathcal{A}_p^{(m)},
    \label{eq:parseval}
\end{equation}
where
\begin{equation}
    \delta\mathcal{A}_p^{(m)} = \vmean{B_p}\cdot\rot\left(\mathcal{E}_\varphi^{(m)}\mathbf{e_\varphi}\right).
    \label{eq:Ap_mode}
\end{equation}

\begin{figure}
	\includegraphics[width=0.99\columnwidth]{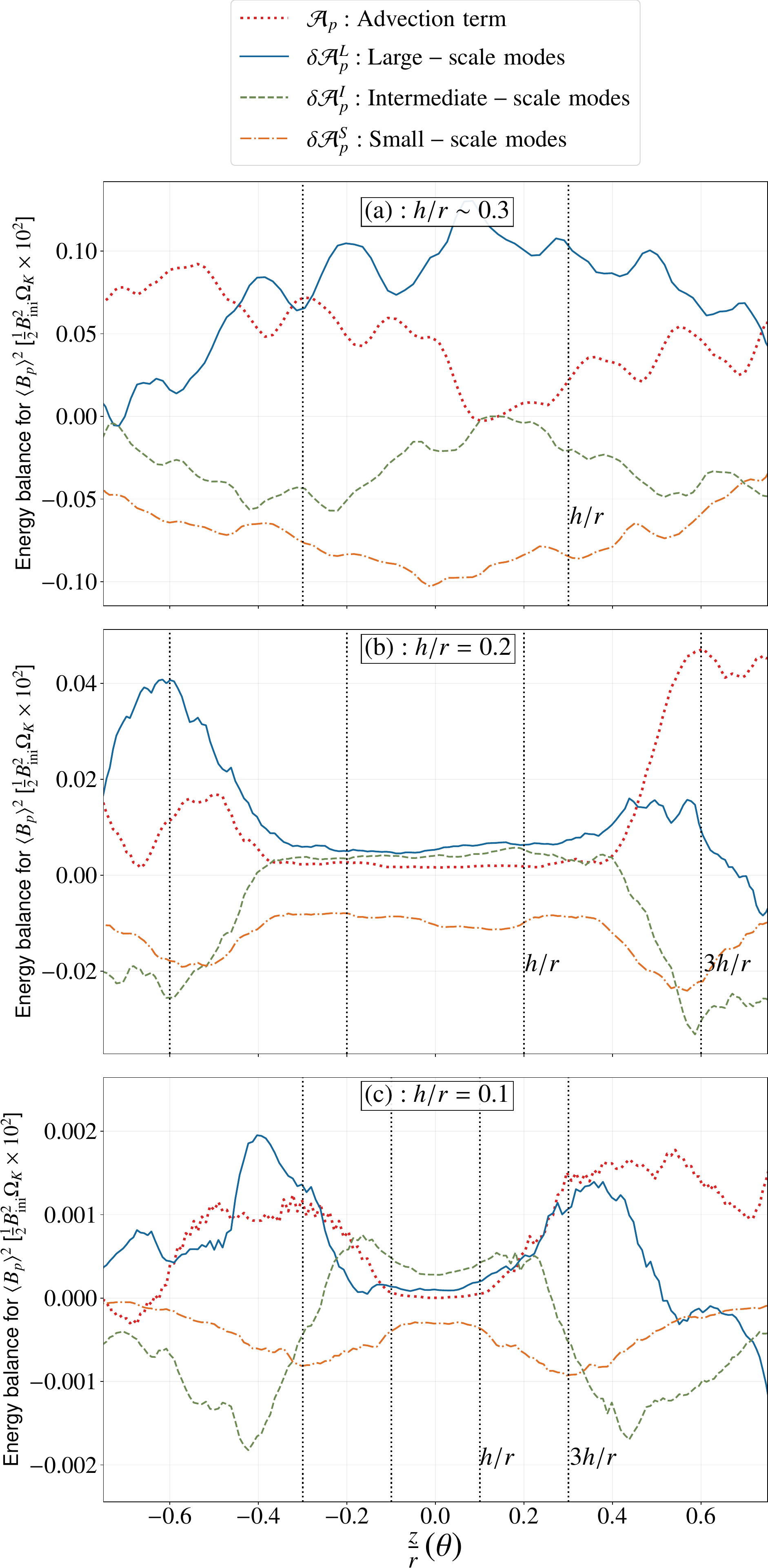}
    \caption{Temporally averaged vertical profiles of the large-, intermediate-, and small-scale contributions to poloidal-field generation, $\delta\mathcal{A}_p^{L}$, $\delta\mathcal{A}_p^{I}$, and $\delta\mathcal{A}_p^{S}$, together with the axisymmetric term $\mathcal{A}_p$, at $r=20r_g$ for $h/r\sim0.3$ (a), $0.2$ (b), and $0.1$ (c). Large-scale modes generate poloidal field at all heights, small-scale modes dissipate it, and intermediate modes generate field near the midplane but dissipate it in the atmosphere. As the disk becomes thinner, both poloidal-field generation and transport become increasingly concentrated at $|z|\sim2$--$4h$.}
    \label{fig:mode_profiles}
\end{figure} 

In Fig.~\ref{fig:poloidal gen} we show $\delta\mathcal{A}_p^{(m)}$ as a function of the azimuthal wavelength normalized by the disk scale height,
\begin{equation}
    \frac{\lambda_\varphi}{h} = \frac{2\pi r}{mh},
    \label{eq:azimuth_wavelenght}
\end{equation}
where $m$ is the azimuthal Fourier mode. The spectrum is normalized by
\begin{equation}
    \delta\mathcal{A}_p^{\rm Norm} =
    \sqrt{\sum\limits_{m=1}^\infty
    |\delta\mathcal{A}_p^{(m)}|^2}.
\end{equation}

Both $\delta\mathcal{A}_p^{(m)}$ and its normalization are evaluated at $r=20r_g$, averaged over time between $t_1$ and $t_2$ (see Table \ref{tab:simulations}), and then averaged over different heights of the disk. We have verified that similar trends are found at other radii, $15r_g\leq r\leq25r_g$. The choice of these heights is motivated by Fig.~\ref{fig:toroidal_gen}. We showed that toroidal field generation is confined to $|\theta-\pi/2|\leq\pi/4$, corresponding to $|z/r|\lesssim0.7$, so we restrict our analysis to this region. Moreover, large scale field generation is subdominant near the disk midplane, for $|z/r|\lesssim h/r$.

These considerations define the three vertical regions shown in Fig.~\ref{fig:poloidal gen}. Panel (a) shows the total contribution, averaged over $|z/r|\lesssim0.7$. Panel (b) shows the disk contribution, averaged over $|z/r|\leq h/r$. Finally, panel (c) shows the disk atmosphere contribution, averaged over $h/r\leq|z/r|\leq0.7$.

In all three panels of Fig.~\ref{fig:poloidal gen} we find the same overall trend: large azimuthal wavelengths contribute positively to poloidal field generation, whereas short wavelengths contribute negatively, acting instead to dissipate the large-scale poloidal field. 

{Beyond the transition from generation to dissipation, the short-wavelength contribution is negative, while its magnitude, $|\delta\mathcal{A}_p^{(m)}|$, decreases approximately as a power law toward smaller $\lambda_\varphi/h$ (equivalently, toward larger $m$). Despite the different disk thicknesses, both the transition wavelength and the shape of this dissipative branch nearly overlap when expressed in terms of $\lambda_\varphi/h$, rather than $m$. This approximate self-similarity suggests that the characteristic scales of poloidal-field generation and dissipation are set primarily by the local disk scale height.}

The critical wavelength separating field generation from dissipation depends, however, on the vertical averaging region. For the total contribution ($|z/r|\lesssim0.7$), the transition occurs at $\lambda_\varphi/h\simeq7$. Restricting the average to the disk ($|z/r|\lesssim h/r$) shifts the transition to much smaller scales, $\lambda_\varphi/h\simeq1.5$--$3$, whereas averaging over the disk atmosphere ($h/r\leq|z/r|\leq0.7$) yields a larger transition scale of $\lambda_\varphi/h\simeq8$--$10$.

Looking only at thick disks, \cite{jacquemin-ide_magnetorotational_2024}  divided the spectrum into two classes: large-scale modes, which generate poloidal magnetic field, and small-scale modes, which dissipate it. While this classification remains a reasonable approximation for the thickest disk, it breaks down for the thinner disks ($h/r\leq0.2)$. In particular, Fig.~\ref{fig:poloidal gen}(b) shows that several intermediate-scale modes contribute positively to field generation within the disk, even though the same modes contribute negatively in the disk atmosphere. Their net contribution therefore depends on the vertical region considered.

Motivated by these results, we divide the spectrum into three regimes. (1) Large-scale modes, which generate poloidal magnetic field both within the disk and in the atmosphere. (2) Intermediate-scale modes, which generate field within the disk but dissipate it in the atmosphere. (3) Small-scale modes, which dissipate the poloidal field in both regions. This naturally defines two characteristic wavelengths: $\lambda_I$, marking the transition from large- to intermediate-scale modes, and $\lambda_S$, marking the transition from intermediate- to small-scale modes.

When normalized to $h$, these characteristic wavelengths  appear to be roughly the same in all of  our simulations, with $\lambda_I/h$  decreasing from $10$ to  $8$ and $\lambda_S/h$ increasing from $1.5$ to $2.5$ as $h/r$ increases from $0.1$ to $0.3$. However, we note that for the thickest disk simulation, modes which contribute negatively to both disk and atmosphere regions exist between $\lambda_S/h$ and $\lambda_I/h$, which might make our mode classification scheme slightly less accurate for that case \citep[see][classification]{jacquemin-ide_magnetorotational_2024}
\begin{figure*}
	\includegraphics[width=0.99\textwidth]{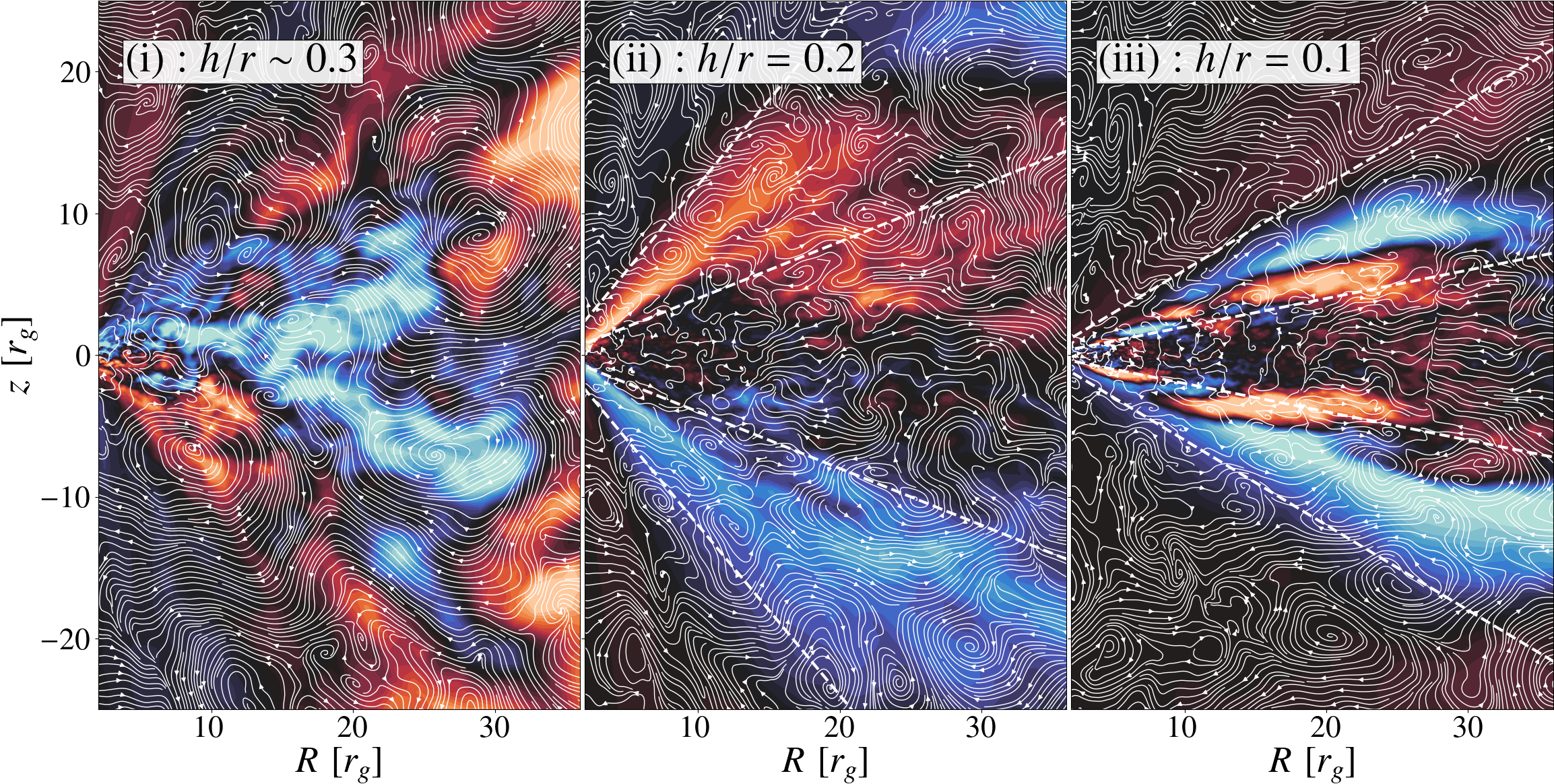}
    \caption{Poloidal velocity structure of the $m=2$ mode for $h/r\sim0.3$ (i), $0.2$ (ii), and $0.1$ (iii). The mode forms coherent rolling motions that preferentially occupy regions of strong toroidal magnetic field. In the thickest disk these rolls are distributed throughout the disk body, whereas in the thinner disks they are concentrated mainly in the atmosphere, at $|z|\sim2$--$5h$ (white dashed lines), consistent with the increasing vertical localization of large-scale poloidal-field generation.}
    \label{fig:stream_mode}
\end{figure*}

Using this classification, we define the contributions of the different mode ranges to poloidal field generation as
\begin{align}
    \delta \mathcal{A}_p^{L} &= \sum\limits_{m=1}^{m_I}\delta\mathcal{A}_p^{(m)},\label{eq:L_modes}\\
    \delta \mathcal{A}_p^{I} &= \sum\limits_{m=m_I+1}^{m_S}\delta\mathcal{A}_p^{(m)},\label{eq:I_modes}\\
    \delta \mathcal{A}_p^{S} &= \sum\limits_{m=m_S+1}^{\infty}\delta\mathcal{A}_p^{(m)},
    \label{eq:S_modes}
\end{align}
where the characteristic mode numbers are defined by
\begin{equation}
    m_{I,S}=2\pi\frac{r}{h}\frac{h}{\lambda_{I,S}}.
\end{equation}

In Fig.~\ref{fig:mode_profiles} we show the vertical profiles of the large-scale, intermediate-scale, and small-scale contributions, $\delta \mathcal{A}_p^{L}$, $\delta \mathcal{A}_p^{I}$, and $\delta \mathcal{A}_p^{S}$, together with the axisymetric contribution $\mathcal{A}_p$. The quantities defined in Eqs.~(\ref{eq:advec_pol}, \ref{eq:L_modes}, \ref{eq:I_modes}, \ref{eq:S_modes}) are plotted as functions of $z/r(\theta)$, normalized by the initial toroidal magnetic energy and the local Keplerian shear rate, for the three disk thicknesses: (a) $h/r\sim0.3$, (b) $h/r=0.2$, and (c) $h/r=0.1$. All quantities are time-averaged between $t_1$ and $t_2$ (see Table \ref{tab:simulations}) and evaluated at $r=20r_g$. We have verified that similar trends are found at other radii, $5r_g\leq r\leq35r_g$.

For all simulations, the large-scale modes contribute positively to poloidal field generation at all heights, while the small-scale modes contribute negatively throughout the system. We find that the large scale non-axisymmetric modes contribute as much as the axisymmetric induction term to the axisymmetric field generation.

{The axisymmetric induction term is always positive independent of disk thickness, this terms can be split into its radial $\mathcal{A}_r$ and latitudinal $\mathcal{A}_\theta$ parts. We start by analyzing its radial part first}

{\scriptsize
\begin{equation}
\begin{aligned}
    \mathcal{A}_r
    ={}&
    \underbrace{-\mean{B_r}\mean{u_r}\,\partial_r\mean{B_r}}_{\text{radial advection}}
    \underbrace{-\frac{\mean{B_r}\mean{u_\theta}}{r}\,\partial_\theta\mean{B_r}}_{\text{latitudinal advection}}
    +\underbrace{\mean{B_r}^{\,2}\,\partial_r\mean{u_r}}_{\text{radial stretching}}
    \\
    &+
    \underbrace{\frac{\mean{B_r}\mean{B_\theta}}{r}\,\partial_\theta\mean{u_r}}_{\text{latitudinal stretching}}
    \underbrace{-\mean{B_r}^{\,2}\,\nabla\cdot\mean{\mathbf{u}_p}}_{\text{compression}} .
\end{aligned}
\label{eq:advec_rad}
\end{equation}
}

{In all three simulations, the largest contributions to $\mathcal{A}_r$ largely cancel, leaving a smaller positive residual. This net amplification is primarily driven by latitudinal stretching in the two thicker disks, consistent with channel-like MRI motions \citep{latter_mri_2010}, and by compression in the thinnest disk, where latitudinal stretching also contributes significantly.}

The latitudinal contribution to axisymmetric induction is subdominant to the radial one, $\mathcal{A}_r\gg\mathcal{A}_\theta$. In the two thinner disks, $\mathcal{A}_\theta$ is dominated by radial advection, as the other contributions largely cancel, giving
\[
\mathcal{A}_\theta \simeq -\vmean{B_\theta}\vmean{u_r}\,\partial_r\vmean{B_\theta}.
\]
In the thickest disk, the balance is less clear, although radial advection remains the largest individual contribution. We examine magnetic-flux transport in more detail in Section~\ref{sec:field_transport}.

The vertical profiles for all the terms differ markedly between the thickest simulation ($h/r\sim0.3$) and the two thinner cases ($h/r\leq0.2$). In the thickest disk, all contributions remain relatively uniform with height. In contrast, the thinner disks exhibit strongly stratified profiles: the different terms are weakest near the disk midplane, $|z/r|\lesssim (1-2) h/r$, and peak at approximately $|z/r|\simeq3h/r$. This stratification is most pronounced for the thinnest disk, where the large-scale modes contribute very little within the body of the disk but become one of the dominant sources of poloidal magnetic energy in the disk atmosphere. {This stratification of the field generation is probably related to the formation of magnetically elevated region as shown in section \ref{sec:glob_stru}.} 

The intermediate-scale modes behave differently. They contribute positively within the disk ($|z/r|\lesssim h/r$) but negatively in the disk atmosphere ($|z/r|\gtrsim3h/r$). This behavior is most pronounced for the thinnest disk ($h/r=0.1$), where the intermediate modes provide the dominant positive contribution to poloidal field generation near the midplane. Their positive contribution decreases steadily with increasing disk thickness and is nearly absent in the thickest simulation. Nevertheless, even in the thickest disk these modes remain one of the dominant sources of poloidal field dissipation in the disk atmosphere.

We conclude that, as the disk becomes thinner, both the generation and transport of the mean poloidal magnetic field become increasingly stratified, occurring primarily in layers located at $2h/r\lesssim|z/r|\lesssim4h/r$. Furthermore, a progressively larger fraction of the field generation is carried by intermediate-scale modes rather than large-scale modes. We speculate that this shift may reduce the efficiency of large-scale field generation, since intermediate-scale modes produce magnetic structures of smaller spatial extent. The combination of these two effects may explain why both the strength of the mean poloidal field and the turbulence decrease as the disk becomes thinner.

Our simulations show that both nonlinear transfer from non-axisymmetric fluctuations, as observed in zero-net-flux shearing boxes \citep{lesur_self-sustained_2008,herault_periodic_2011,riols_global_2013,mamatsashvili_zero_2020}, and direct induction by axisymmetric MRI motions contribute to the generation of large-scale magnetic fields. These mechanisms are not mutually exclusive, as stratified local simulations with weak net vertical flux also exhibit large-scale dynamo activity and magnetic-field reversals \citep{salvesen_accretion_2016}. Our results suggest that both processes operate in global disks with comparable importance.

\subsection{Mode structure}
It is clear from the previous section that the large-scale modes play a central role in the generation of the mean poloidal magnetic field. To better understand their structure, we examine the poloidal velocity field of the $m=2$ Fourier mode. Specifically, we plot the streamlines of $\mathbf{u}_p^{m=2}e^{2i\varphi}$ evaluated at $\varphi=0$, superimposed on the $\varphi$-averaged toroidal magnetic field. The resulting streamlines are shown in Fig.~\ref{fig:stream_mode} for the three disk thicknesses: (i) $h/r\sim0.3$, (ii) $h/r=0.2$, and (iii) $h/r=0.1$, evaluated at $t=5.5\times10^3,r_g/c$, $3.85\times10^4,r_g/c$, and $3.9\times10^4,r_g/c$, respectively. For the two thinner simulations ($h/r\leq0.2$), we also indicate the heights $|z|=2h$ and $|z|=5h$.

In all three simulations, the $m=2$ mode consists of coherent rolling structures in the poloidal plane. These should not be confused with axisymmetric convective eddies, as they possess an $m=2$ azimuthal dependence. In every case, the rolls are predominantly located in regions where the toroidal magnetic field is strongest, suggesting a close connection between the two.

The vertical distribution of these structures, however, depends strongly on the disk thickness. In the thickest disk ($h/r\sim0.3$), the rolls are distributed throughout the disk body with no preferred height. In contrast, for the thinner disks ($h/r\leq0.2$), they are concentrated almost exclusively within the disk atmosphere, between $|z|\simeq2h$ and $|z|\simeq5h$, as indicated by the white dashed lines in Fig.~\ref{fig:stream_mode}. Below $|z|\simeq2h$, the flow no longer exhibits coherent rolling motions, but instead appears significantly more disordered.

This behavior is fully consistent with the results of the previous section. In the thinner disks, the atmospheric region is where large-scale poloidal field generation is strongest, whereas in the thickest disk, field generation is distributed much more uniformly throughout the disk. The vertical localization of these coherent rolls therefore appears to be closely linked to the stratification of the dynamo itself. More generally, the figure reveals a trend toward increasingly local MRI behavior as the disk becomes thinner, transitioning from a relatively global, large-scale MRI in the thickest disk to a more localized dynamo concentrated within a narrower region of the disk.

We note that \cite{lesur_localized_2008} showed that non-axisymmetric MRI modes attain their maximum growth rate near the maxima of the toroidal magnetic field when the latter exhibits a complex vertical structure. The same study also found that, depending on the absolute strength of the toroidal field, the opposite behavior can occur, with the fastest-growing modes avoiding the field maxima. We hypothesize that the former regime may be related to the behavior observed here, although a more detailed investigation is needed to establish this connection.

\section{Large scale transport}
\label{sec:transport}
\subsection{Angular momentum transport}
\label{sec:ang_transport}

\begin{figure}
	\includegraphics[width=\columnwidth]{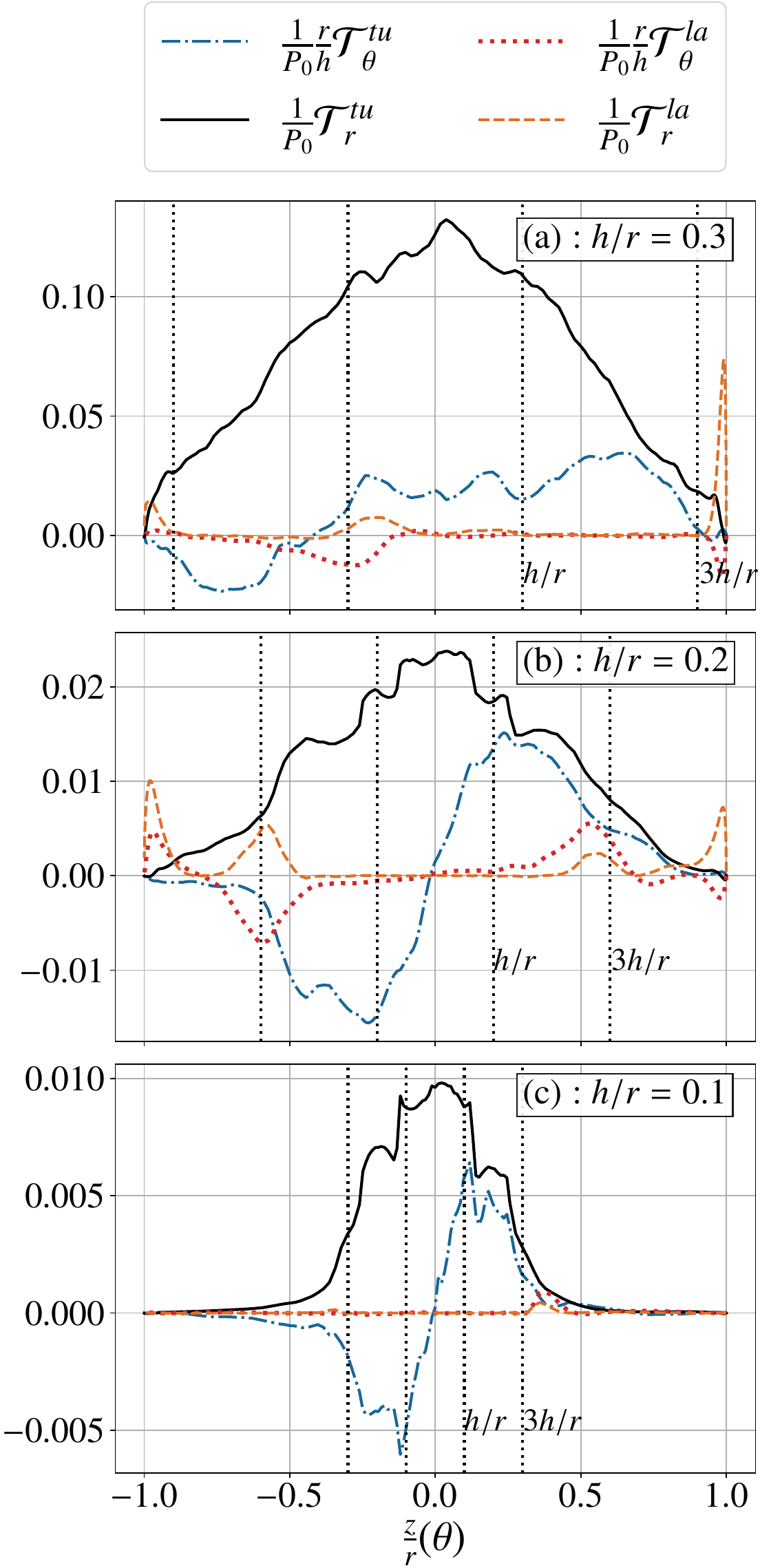}
    \caption{Temporally averaged vertical profiles of the radial and latitudinal laminar and turbulent stresses at $r=20r_g$ for $h/r\sim0.3$ (a), $0.2$ (b), and $0.1$ (c), normalized to the midplane pressure. The latitudinal stresses are additionally normalized by $h/r$ to account for their different contribution to the accretion velocity. The radial turbulent stress is the largest individual component in all simulations, while for $h/r\leq0.2$ the latitudinal turbulent stress contributes comparably to angular-momentum transport.}
    \label{fig:Stresses}
\end{figure} 

We have seen in Section \ref{sec:turbulent_behavior} that $\alpha_{\nu}$ decreases with decreasing disk thickness.  Since this quantity drives angular momentum transport and therefore the accretion rate, decreasing it will also affect the velocity at which gas and the magnetic field are transported inward toward the black hole. To understand if this is the case we investigate in detail the angular momentum transport in our disk models.  Accretion disks are subject to laminar (mean) and turbulent stresses
\begin{align}
    \mathbf{\mathcal{T}^{la}} &= -\frac{1}{4\pi}\vmean{B_p}\mean{B_\varphi} \label{eq:stress_la}\\
    \mathbf{\mathcal{T}^{tu}} &= \mean{\rho}\mean{\delta u_p\delta u_\varphi}-\frac{1}{4\pi}\mean{\delta\mathbf{B_p}\delta B_\varphi}, \label{eq:stress_tu}
\end{align}
which can be further decomposed into latitudinal and radial components. In most works one only considers the radial turbulent stress, which is related to the \cite{shakura_black_1973} $\alpha_\nu$ parameter by
\begin{equation}
    \alpha_{\nu} = \frac{1}{P_0}\mathcal{T}^{tu}_{r}.
\end{equation}
However, in an accretion disk with large scale vertical magnetic fields and vertical (latitudinal) turbulent torques, all terms in Eqs.~(\ref{eq:stress_la}, \ref{eq:stress_tu}) should be considered. Therefore we also define the total stress
\begin{equation}
    \mathbf{\mathcal{T}} = \mathbf{\mathcal{T}^{la}}+\mathbf{\mathcal{T}^{tu}}.
\end{equation}

Once all stress components are known, an approximate expression for the accretion velocity can be obtained \footnote{Strictly speaking, this expression should be written in terms of the cylindrical components of the stresses. Nevertheless, we find that the spherical form provides a good approximation to the accretion velocity.} \citep{casse_magnetized_2000,lesur_systematic_2021,begelman_simple_2024},
\begin{equation}
\label{eq:vacc_theory}
v_{\rm acc}^{\rm th} \simeq -\frac{2V_k}{P_0}\frac{h}{r}\left(\frac{h}{r}\mathcal{T}_{\rm r}|_{z=0}+\mathcal{T}_{\theta}|_{z=h}-\mathcal{T}_{\theta}|_{z=-h}\right),
\end{equation}
where we evaluate the radial stress at $z=0$ and the vertical stresses at the disk surface $z=\pm h$. Equation~\ref{eq:vacc_theory} illustrates why surface stresses become increasingly important in thinner disks: their contribution to the accretion velocity scales linearly with $h/r$, whereas the contribution from radial stresses scales quadratically.

In Fig.~\ref{fig:Stresses} we show the different stress terms defined in Eqs.~(\ref{eq:stress_la}, \ref{eq:stress_tu}) which include all components of angular momentum transport leading to accretion as function of $\theta$ and normalized to the midplane pressure $P_0$ at $r=20r_g$ for the three disk thicknesses: $\mathrm{(a):}~h/r\sim0.3$, $\mathrm{(b):}~h/r=0.2$, and $\mathrm{(c):} ~h/r=0.1$. The stresses and pressures are time-averaged between $t_1$ and $t_2$ (see Table \ref{tab:simulations}). We normalize the latitudinal stresses, $\mathcal{T}^{\rm la}_{\theta}$ and $\mathcal{T}^{\rm tu}_{\theta}$, by $h/r$ because their contribution to the radial accretion velocity scales differently from that of the radial stresses (see Eq.~\ref{eq:vacc_theory}). Specifically, radial stresses contribute as $(h/r)^2$, whereas latitudinal stresses contribute as $h/r$, since they act on the disk surface rather than throughout its volume. We find very similar trends at all radii examined, $5r_g\leq r\leq30r_g$.

For all disk thicknesses considered, the largest stress component is the radial turbulent stress, $\mathcal{T}^{\rm tu}_{r}$. However, this does not necessarily imply that it is the dominant driver of accretion. This stress follows a Gaussian-like profile, with angular momentum transport concentrated near the disk midplane. As expected, the profile becomes progressively thinner as $h/r$ decreases.

After accounting for their different scaling with the accretion velocity, we find that the latitudinal turbulent stresses, $\mathcal{T}^{\rm tu}_{\theta}$, contribute approximately as much to angular momentum transport as the radial turbulent stress for the $h/r\leq0.2$ simulations. In contrast, for the thickest disk ($h/r\sim0.3$), the radial turbulent stress remains the dominant contributor to accretion.

\begin{figure}
	\includegraphics[width=\linewidth]{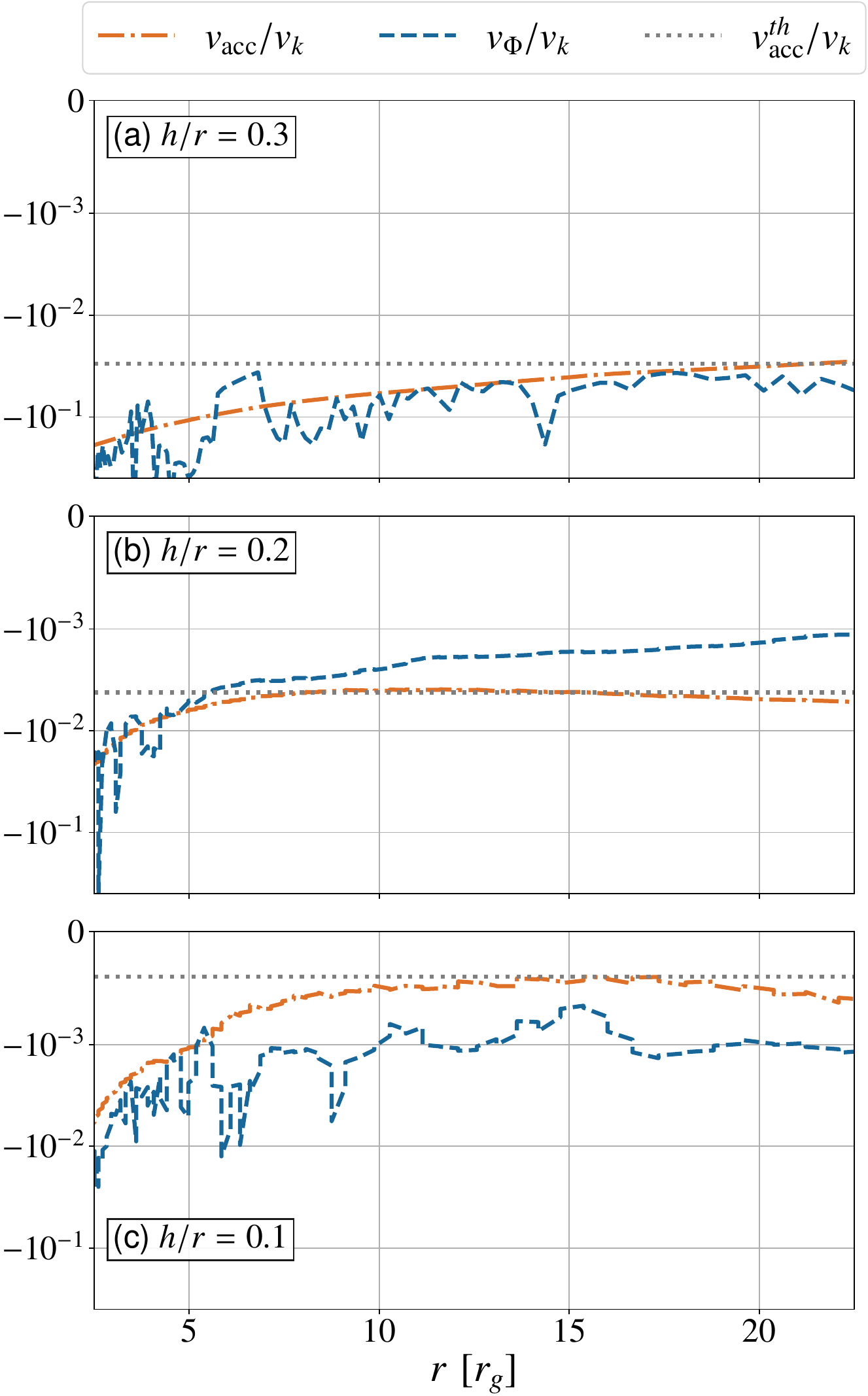}
    \caption{Radial profiles of the temporally averaged accretion velocity, $v_{\rm acc}$, and magnetic-flux transport velocity, $v_\Phi$, normalized by the Keplerian velocity, for $h/r\sim0.3$ (a), $0.2$ (b), and $0.1$ (c). The dotted horizontal lines show the prediction from the angular-momentum transport model, which agrees well with the measured $v_{\rm acc}$ at $r\gtrsim10r_g$. The flux transport velocity remains comparable to the accretion velocity in all simulations, and both decrease steeply as the disk becomes thinner.}
    \label{fig:accretion_vel}
\end{figure} 

For all simulations, the laminar (mean) stresses, $\mathbf{\mathcal{T}^{\rm la}}$, are negligible and can safely be ignored. This is expected because the mean poloidal magnetic energy, $\mean{B_p}^2$, is at least an order of magnitude smaller than the mean toroidal magnetic energy, $\mean{B_\varphi}^2$, which is itself roughly an order of magnitude weaker than the turbulent magnetic energy, $\mean{\delta B^2}$. Consequently, the laminar stresses, which arise from products of these already weak mean fields, remain dynamically unimportant. This situation changes  when a large-scale magnetic field becomes sufficiently strong for laminar stresses to remove angular momentum as efficiently as the turbulent stresses \citep{mishra_strongly_2020,jacquemin-ide_magnetic_2021,manikantan_winds_2024,scepi_magnetic_2024}. We therefore neglect the laminar stresses throughout the remainder of this manuscript.

To verify that all relevant angular momentum transport processes are properly accounted for, we compute the accretion velocity as
\begin{equation}
v_{\rm acc}=\frac{1}{\int\limits_0^\pi\mean{\rho}\mathrm{d}\theta}
\int\limits_0^\pi\mean{\rho u_r}\mathrm{d}\theta,
\label{eq:vacc}
\end{equation}
which is just the radial velocity weighted by the density.

The resulting accretion velocity, normalized by the Keplerian velocity, is shown as the dashed blue line in Fig.~\ref{fig:accretion_vel}(a--c) as a function of radius for the three disk thicknesses: (a) $h/r\sim0.3$, (b) $h/r=0.2$, and (c) $h/r=0.1$. We have time-averaged the accretion velocity between $t_1$ and $t_2$ (see Table \ref{tab:simulations}).
As the disk becomes thinner, the accretion velocity decreases dramatically, from $4\times10^{-2}V_k$ for $h/r\sim0.3$, to $4.4\times10^{-3}V_k$ for $h/r=0.2$, and finally to $4.5\times10^{-4}V_k$ for $h/r=0.1$. Figure~\ref{fig:accretion_vel} also shows the prediction of Eq.~\ref{eq:vacc_theory}, computed using the stresses measured in Fig.~\ref{fig:Stresses}, as the dotted gray line. 

We find good agreement between the measured accretion velocity and the prediction of Eq.~\ref{eq:vacc_theory} (the grey dotted line) for $r\gtrsim10r_g$, with the agreement being particularly good for the thinner disks ($h/r\leq0.2$). The largest deviations occur for the thickest disk, which is unsurprising since Eq.~\ref{eq:vacc_theory} is derived under the thin-disk approximation as a generalization of the \cite{shakura_black_1973} model. In geometrically thick disks, the accretion velocity is no longer expected to scale simply with the Keplerian velocity. Nevertheless, the agreement is sufficiently good to conclude that the dominant mechanisms driving accretion in our simulations are well understood. {The agreement is also bad close to to the black hole, $r\leq8r_g$ , as the torques and accretion velocity grow considerably there, do to horizon effects \citep{penna_shakura-sunyaev_2013}}

The theory of \cite{shakura_black_1973} predicts an accretion velocity scaling of $v_{\rm acc}^{\rm SS}\propto\alpha(h/r)^2$, whereas we find a much steeper relation, $v_{\rm acc}\propto(h/r)^4$. This difference arises for two reasons. First, unlike the standard $\alpha$-disk model, the turbulence strength is itself a strong function of disk thickness. In Section~\ref{sec:turbulent_behavior} we found $\alpha_\nu\propto(h/r)^{2.5}$, likely because the efficiency of magnetic field generation also decreases with $h/r$. Second, our model includes turbulent vertical torques, which are neglected in the \cite{shakura_black_1973} theory. Accounting for these torques yields a more complete description of angular momentum transport and roughly reproduces the observed scaling of the accretion velocity. 

The radial velocity drives the large-scale transport of both mass and magnetic flux. Having established the origin of the accretion flow, we now examine how it governs the large-scale transport of magnetic flux.

\subsection{Large scale magnetic field transport}
\label{sec:field_transport}


To understand why the magnetic flux accumulates onto the black hole at such different rates, we must examine how magnetic flux is transported through the accretion disk. We compute the magnetic flux threading the northern hemisphere as
\begin{equation}
    \Phi(r,t) = 2\pi \int\limits_0^{\pi/2}\mean{ B^{r}} \sqrt{-g}\diff \theta ,
    \label{Eq:pol_flux}
\end{equation} 
where we do not take the absolute value in order to retain information about magnetic polarity reversals.
\begin{figure*}
	\includegraphics[width=\textwidth]{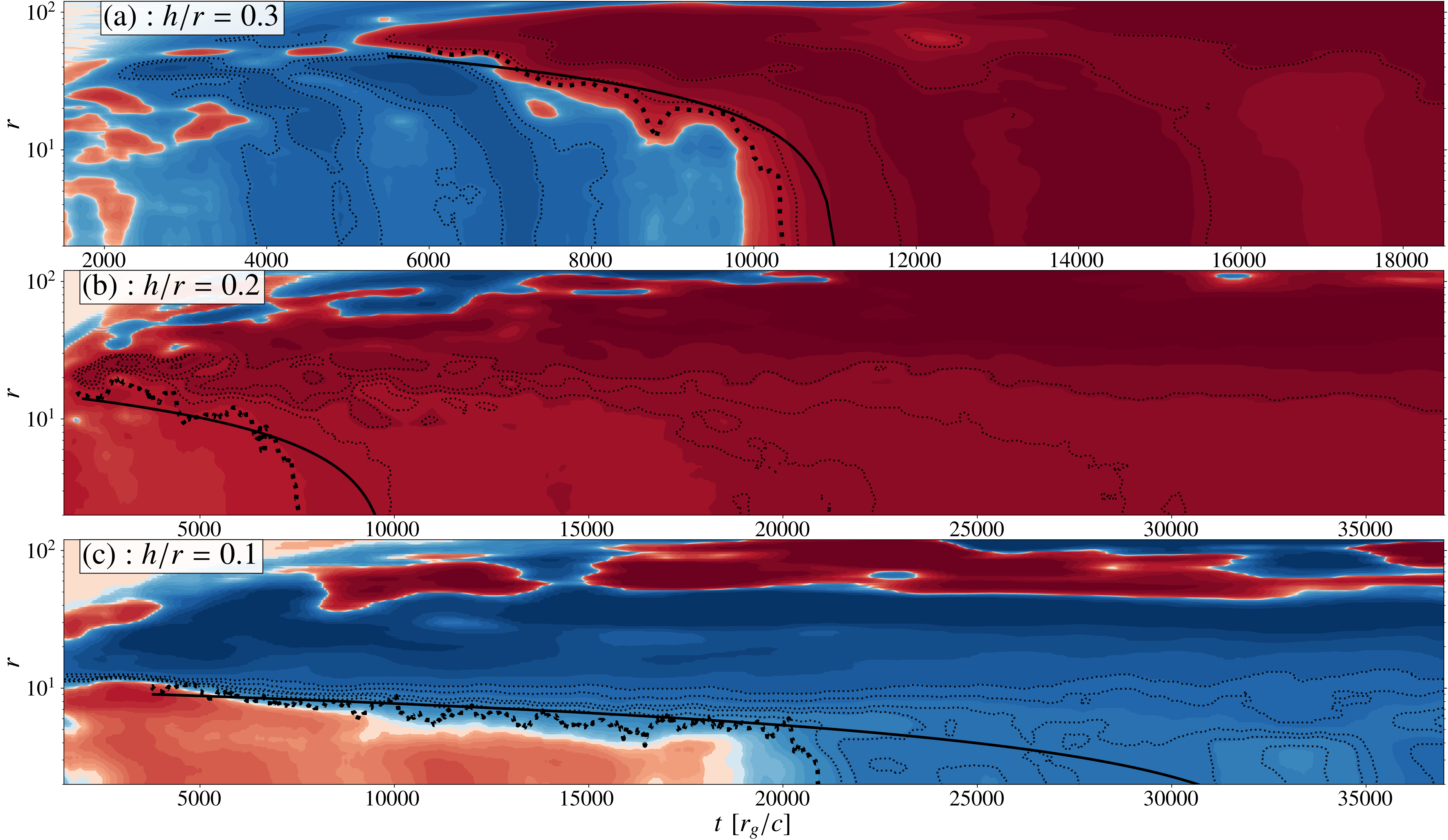}
    \caption{Spacetime evolution of the signed magnetic flux, $\Phi(r,t)$, for $h/r=0.3$ (a), $0.2$ (b), and $0.1$ (c). Colours indicate magnetic polarity and contours trace magnetic flux surfaces as they evolve through the disk. In all simulations, progressively larger magnetic loops are generated and advected inward, producing a coherent large-scale flux structure at late times. The dotted black curves traces representative magnetic flux surfaces identified in the simulation, while the solid black lines show a corresponding analytical trajectories. The agreement deteriorates at small radii ($r\leq5r_g$), where ISCO and near-horizon dynamics are not captured by the simple transport model. The markedly slower transport in thinner disks explains their correspondingly slower black-hole flux accumulation.}
    \label{fig:triple_phi_rt}
\end{figure*} 

Figure~\ref{fig:triple_phi_rt} shows an $r$--$t$ spacetime diagram of the magnetic flux threading the disk midplane for our three simulations with different disk thicknesses: (a) $h/r=0.3$, (b) $h/r=0.2$, and (c) $h/r=0.1$. Different colors denote opposite magnetic polarities, while the contours trace magnetic flux surfaces as they evolve through the disk.

In all three simulations, magnetic field loops are generated within the disk and subsequently advected inward. As the simulations evolve, progressively larger loops are produced and transported toward the black hole, increasing the accumulated magnetic flux. By $t=4\times10^4r_g/c$, all simulations have developed a coherent large-scale magnetic structure of a single polarity inside $r\lesssim40r_g$. However, the strength of this large-scale field differs dramatically between the simulations, as we discuss in Section~\ref{sec:turbulent_behavior}. In the thinnest disk, this inward transport becomes less apparent in Fig.~\ref{fig:triple_phi_rt} for $t\gtrsim2.5\times10^4r_g/c$ and $r\gtrsim10r_g$, where some flux surfaces appear to remain at approximately fixed radii.

Using the formalism of \cite{jacquemin-ide_demystifying_2026}, we estimate the magnetic flux transport velocity as
\begin{equation}
    v_\Phi = -\partial_t\Phi/\partial_r\Phi,
\end{equation}
where $\Phi$ is first smoothed in time over a window of 50 dumps and then averaged between $t_1$ and $t_2$ (see Table \ref{tab:simulations}). Because this quantity is particularly noisy, especially for the thinner disks where the magnetic flux is weak, we average only the negative values of $v_\Phi$, corresponding to inward transport. This procedure is motivated by the fact that magnetic flux follows complex zigzag trajectories, repeatedly advancing and retreating, making numerical differentiation particularly noisy. However, this negative average is verified below by comparing to Fig.~\ref{fig:triple_phi_rt}.

The resulting transport velocity, normalized by the Keplerian velocity, is shown in Fig.~\ref{fig:accretion_vel}. We find that $v_\Phi$ is comparable to the accretion velocity, exhibiting a similar radial profile and nearly the same dependence on disk thickness. As the disk becomes thinner, the transport velocity decreases rapidly. For the thickest disk, $v_\Phi$ is almost identical to $v_{\rm acc}$. For the intermediate-thickness disk, it is smaller by roughly a factor of $2$--$3$, while for the thinnest disk it exceeds the accretion velocity by approximately a factor of $2$. These results are consistent with the prediction of \cite{jacquemin-ide_demystifying_2026} that, outside the MAD state, the magnetic flux transport velocity should be comparable to the accretion velocity, despite the very different magnetic field configurations considered here.

{To verify our estimate of $v_\Phi$, the dotted black curve in Fig.~\ref{fig:triple_phi_rt} traces a representative magnetic flux surface identified in the simulation. To facilitate comparison with the measured transport velocity, we approximate its radial dependence as Keplerian, $v_\Phi\propto r^{-1/2}$, although this scaling is not exact throughout the flow. We can then compare the simulated flux surface with the analytical trajectory obtained by integrating}
\begin{equation}
    \dot{R}_\phi =- \hat{v}\Tfrac{r}{r_g}^{-1/2},
\end{equation}
which gives
\begin{equation}
    R_\phi = \left(-\frac{3}{2}\hat{v}(t-t_i)+r_i^{3/2}\right)^{2/3},
\end{equation}
where $r_i$ and $t_i$ denote the initial position and time of the trajectory. We adopt $\hat{v}=4\times10^{-2}$, $4.4\times10^{-3}$, and $6\times10^{-4}$ for the $h/r=0.3$, $0.2$, and $0.1$ simulations, respectively.

For $r\gtrsim5r_g$ the analytical trajectories reproduce the evolution of the simulated magnetic flux surfaces shown in Fig.~\ref{fig:triple_phi_rt}, demonstrating that our estimate of the magnetic flux transport velocity captures the global evolution of the large-scale field. The reason our analytical estimate breaks down for $r\lesssim5r_g$ is due to the horizon effects leading to a faster velocity than Keplerian near the BH (see Fig.~\ref{fig:accretion_vel}).

The steep decline of $v_\Phi$ with decreasing $h/r$ is therefore one of the primary reasons why magnetic flux accumulates much more slowly onto the black hole in thinner disks (Fig.~\ref{fig:phi_BH} and \cite{penna_shakura-sunyaev_2013}). 

In Appendix~\ref{A:correl_lenght}, we compute the radial correlation length of the magnetic flux loops. We find that their radial extent is consistently of order $h$, helping to explain why magnetic-flux accumulation onto the black hole decreases so rapidly with decreasing disk thickness.

\section{Connection to jet launching and MAD formation}
\label{sec:implications}

In this section, we compile the steep dependencies of the turbulence strength, mean-field strength, magnetic-flux advection velocity, and loop size on $h/r$. We then use these scalings to understand the accumulation of magnetic flux near the black hole and its steep dependence on $h/r$. We then estimate the time required for our simulated disks to reach the MAD state.  

\subsection{Scalings of field generation and transport with disk thickness}
\label{sec:scalings}
All simulations begin with the same toroidal magnetic field strength, but develop strongly different mean fields and turbulent properties as $h/r$ decreases. We evaluate the following scalings at $r=20\,r_g$, sufficiently far from the black hole to limit relativistic effects, but close enough for substantial flux advection. Their dependence on radius and evolutionary stage remains uncertain, as do possible effects of the different hydrodynamic initial conditions (Section~\ref{sec:context_limits}). The resulting exponents should therefore be regarded as approximate.

As shown in Fig.~\ref{fig:betas_time}, we measure $\beta_p^{-1}\simeq3\times10^{-3}$, $10^{-4}$, and $2\times10^{-5}$ for the $h/r=0.3$, $0.2$, and $0.1$ simulations, respectively. These values correspond approximately to a scaling
\begin{equation}
    \beta_p^{-1}\propto(h/r)^{4.5}.
\end{equation}
Furthermore, as shown in Fig.~\ref{fig:lenght_scale} the size of the axisymmetric loops scales roughly as $h$ the disk thermal scale height.

We characterize the strength of the turbulence using the turbulent $\alpha_\nu$ coefficient \citep{shakura_black_1973}. This scaling is a direct consequence of the decreasing strength of the generated magnetic field. As shown in Fig.~\ref{fig:alpha}, we find $\alpha_\nu\simeq1.3\times10^{-1}$, $3\times10^{-2}$, and $8\times10^{-3}$ for the $h/r=0.3$, $0.2$, and $0.1$ simulations, respectively, corresponding to
\begin{equation}
    \alpha_\nu\propto(h/r)^{2.5}.
\end{equation}

{The accretion velocity is expected to decrease with disk thickness because it is set by the efficiency of angular-momentum transport and, even at fixed turbulent stress, depends strongly on $h/r$. The magnetic-flux transport velocity is set by the competition between radial inflow and turbulent magnetic diffusion. As shown in Section~\ref{sec:field_transport}, we find that it remains comparable to the accretion velocity across our simulations and likewise decreases toward thinner disks.} The advection and accretion velocities are shown in Fig.~\ref{fig:accretion_vel}. We find $ v_{\rm acc}/V_k\simeq3\times10^{-2},\,4\times10^{-3},\,4\times10^{-4}
$ and $v_{\phi}/V_k\simeq3\times10^{-2},\,2\times10^{-3},\,6\times10^{-4}
$ for the $h/r=0.3$, $0.2$, and $0.1$ simulations, respectively. This corresponds to an approximate scaling
\begin{equation}
    v_\Phi/V_k\propto v_{\rm acc}/V_k\propto(h/r)^{3.5}.
\end{equation}
{This scaling is steeper than what is naively expected from a thin \cite{shakura_black_1973} like disk ($\Tfrac{h}{r}^{2}$, the reason for this is that the $\alpha_\nu$ also decrease with $h/r$.}

{If we take the measured dependence of $\beta_p^{-1}$ on $h/r$ as given, the remaining scalings follow naturally. Using the MRI relation $\alpha\sim\sqrt{\beta_p^{-1}}$ \citep{salvesen_accretion_2016}, we obtain}
\begin{equation}
\alpha\propto\left(\frac{h}{r}\right)^{2.25},
\end{equation}
{close to the measured $\alpha\propto(h/r)^{2.5}$ dependence. Substituting this into Eq.~\ref{eq:vacc_theory}, the radial and vertical stress contributions scale as}
\begin{align}
\left(\frac{v_{\rm acc}}{V_k}\right){\rm radial}
&\propto \left(\frac{h}{r}\right)^{4.25},\\
\left(\frac{v_{\rm acc}}{V_k}\right){\rm vertical}
&\propto \left(\frac{h}{r}\right)^{3.25}.
\end{align}
{The total accretion velocity is therefore expected to lie between these limits, consistent with the measured $v_{\rm acc}/V_k\propto(h/r)^{3.5}$. This suggests that, once the dynamo sets the mean poloidal field strength, the corresponding trends in turbulent stress, accretion and advection velocities follow approximately from it.}

{Reproducing the dependence of the mean field on $h/r$ is considerably more difficult, since it reflects both changes in the spectral distribution of mean-field generation and changes in the axisymmetric induction terms identified in Section~\ref{sec:field_gen}. }

\subsection{Flux reservoir and MAD accumulation timescale}
\label{sec:time_to_MAD}

\begin{figure}
	\includegraphics[width=\columnwidth]{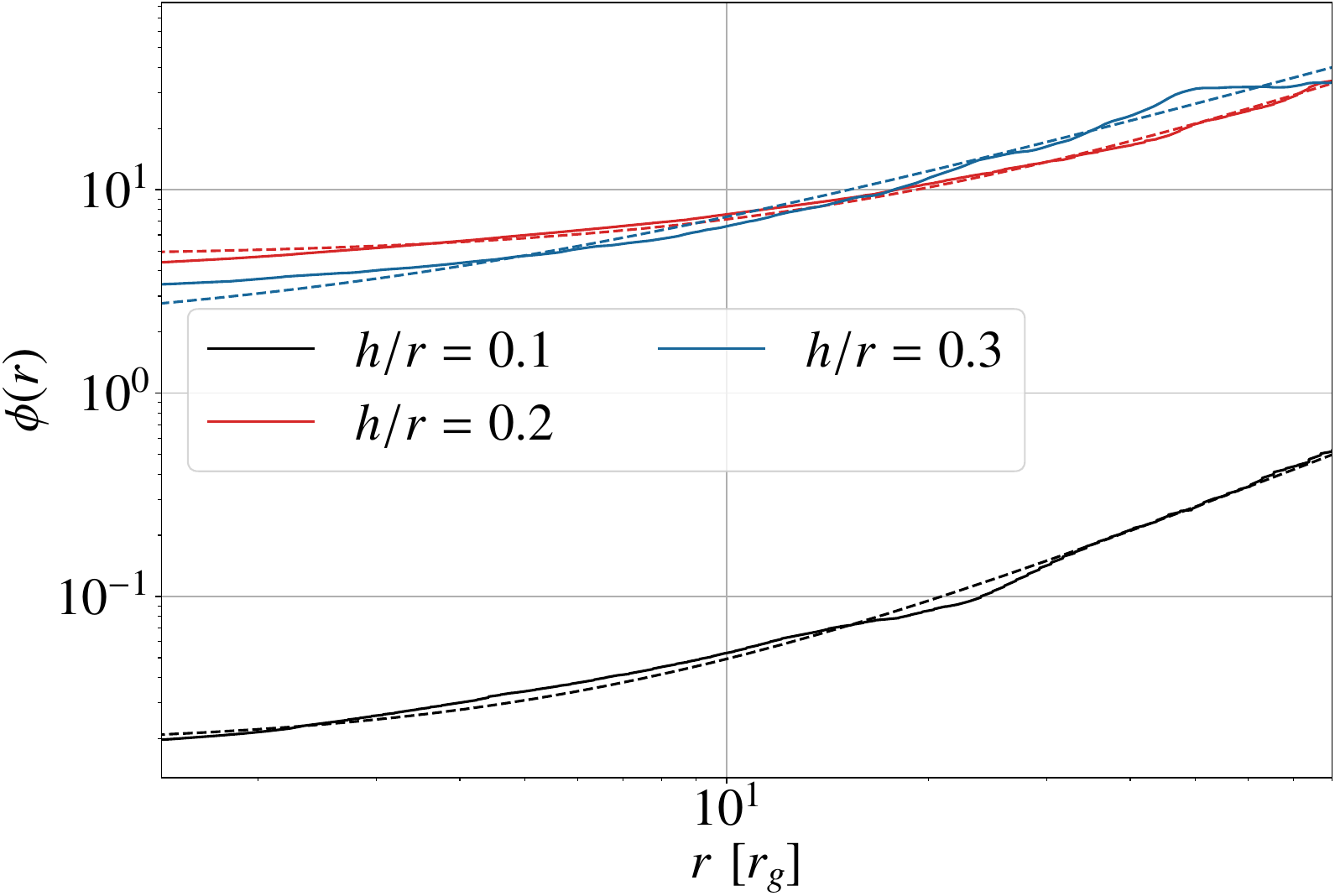}
    \caption{Time-averaged radial profiles of the dimensionless magnetic flux, $\phi(r)$, for the three disk thicknesses. The dashed lines show power-law fits, $\phi(r)=\phi_{\rm BH}^{t_1^2}+a_\phi(r/r_g)^\xi$. All simulations contain substantially more magnetic flux at larger radii than has yet reached the black hole, indicating that the thinner disks could continue to build flux over much longer advection timescales. We caution that the thickest disk is averaged over an earlier evolutionary stage because it reaches MAD much sooner, which reduces its apparent flux signal relative to what might otherwise be expected. The main difference between simulations is the strong decrease in the flux normalization $a_\phi$ with decreasing $h/r$.}
    \label{fig:pow_law}
\end{figure}

To assess whether the thinner disks could eventually reach MAD, we examine the time-averaged radial magnetic-flux profiles $\phi(r)$ in Fig.~\ref{fig:pow_law}, where $\phi(r)$ is defined in Section~\ref{sec:flux_hole} and the averaging interval $[t_1,t_2]$ is given in Table~\ref{tab:simulations}. These profiles are well described by
\begin{equation}
    \phi(r)=\phi_{\rm BH}^{1,2}+a_\phi\left(\frac{r}{r_g}\right)^\xi,
\end{equation}
with $\xi\simeq0.9$, $1.2$, and $1.3$, and $a_\phi\simeq0.7$, $0.15$, and $2\times10^{-3}$ for the three disk thicknesses $h/r=0.3$, $0.2$, and $0.1$, respectively. Thus, while the radial exponent varies relatively little, the flux normalization decreases steeply with $h/r$. We caution that the thickest disk is averaged over an earlier evolutionary stage, before reaching MAD, so the similarity of its radial profile to that of the intermediate disk does not imply comparable flux-accumulation timescales.

The measured radial profiles can be interpreted as a lower bound on the magnetic-flux reservoir, since continued MRI dynamo action may generate and amplify additional large-scale flux. All three simulations already contain substantially more magnetic flux at larger radii than has reached the black hole. In principle, this reservoir could eventually drive the thinner disks toward MAD, provided the flux survives transport over the required distances. However, the corresponding advection timescales are much longer than the simulated evolution. Using $t_{\rm adv}=R/v_\Phi$, we find $t_{\rm adv}\simeq3\times10^5\,r_g/c$ for $h/r=0.2$ and $t_{\rm adv}\simeq1.7\times10^6\,r_g/c$ for $h/r=0.1$ at $R=100r_g$. Thus, the absence of MAD in these simulations does not preclude its eventual formation over much longer timescales.

We now estimate the time required for the thinner disks to reach MAD. Using the fitted radial flux profiles and retaining only the dominant contribution from the radial flux reservoir, the minimum radius enclosing sufficient magnetic flux is
\begin{equation}
    r^{\rm min}_{\rm MAD} \sim \Tfrac{\phi_{\rm MAD}}{a_\phi}^{1/\xi}.
    \label{eq:rminmad}
\end{equation}
Assuming that the flux-advection velocity follows a Keplerian radial scaling,
\begin{equation}
    |v_\Phi| = \hat{v}\Tfrac{r}{r_g}^{-1/2},
\end{equation}
the advection timescale can be estimated as $t_{\rm adv}\sim r/|v_\Phi|$. The time required to advect the necessary flux to the black hole is approximately the advection time evaluated at $r^{\rm min}_{\rm MAD}$. This gives
\begin{equation}
    t_{\rm MAD} \sim \frac{1}{\hat{v}}
    \Tfrac{\phi_{\rm MAD}}{a_\phi}^{\frac{3}{2\xi}}.
    \label{eq:t_MAD}
\end{equation}

For the thickest disk, adopting $\hat{v}=3\times10^{-2}$ and $\phi_{\rm MAD}=50$ gives $t_{\rm MAD}\simeq4\times10^4\,r_g/c$, within a factor of 2 to the MAD transition observed in the simulation. For the intermediate disk, $\hat{v}=2\times10^{-3}$ and $\phi_{\rm MAD}=50$ yield $t_{\rm MAD}\sim7\times10^5\,r_g/c$, potentially accessible to future long-duration simulations. For the thinnest disk, adopting $\hat{v}=4\times10^{-4}$ and $\phi_{\rm MAD}=20$, consistent with recent studies of thin MADs \citep{avara_efficiency_2016,scepi_magnetic_2024,lowell_evidence_2025}, gives $t_{\rm MAD}\sim10^8\,r_g/c$, well beyond currently feasible simulation timescales. Thus, although the thinner disks contain enough magnetic flux at large radii to potentially reach MAD, the combined weakening of the flux reservoir and slowing of its inward transport produce an extremely steep increase in the accumulation timescale as $h/r$ decreases, reaching nearly four orders of magnitude between our thickest and thinnest disks.



As a consistency check, we replace $\phi_{\rm MAD}$ in Eq.~\ref{eq:t_MAD} with the measured horizon fluxes, $\phi_{\rm BH}^{1,2}=5$ and $0.02$ for the intermediate and thinnest disks, respectively. This gives accumulation times of $\sim4.0\times10^4\,r_g/c$ and $\sim3.6\times10^4\,r_g/c$, respectively, differing from the corresponding averaging times by at most $\sim20\%$. Despite its simplicity, our estimate therefore reproduces the characteristic timescale of magnetic-flux accumulation reasonably well.

These estimates assume that the measured flux reservoir and advection velocities remain representative of the disk's subsequent evolution. Continued dynamo action could generate additional magnetic flux, while the accumulation of large-scale fields may modify the turbulent stresses, accretion rate, disk structure, and transport velocity \citep{jacquemin-ide_magnetic_2021,roberts_global_2026}. A self-consistent treatment would therefore require following the coupled evolution of the magnetic field and accretion flow. Our estimates should consequently be interpreted as order-of-magnitude extrapolations rather than detailed predictions of long-term disk evolution.

\subsection{What can we say about jets?}

Reaching MAD is not a prerequisite for launching a detectable Blandford--Znajek jet \citep{blandford_electromagnetic_1977}. We therefore expect the timescale for observable jet formation to be shorter than that required to reach MAD,
\begin{equation}
    t_{\rm jet}\lesssim t_{\rm MAD}.
\end{equation}
However, since the Blandford--Znajek jet power scales approximately as $P_{\rm jet}\propto\Phi_{\rm BH}^2$, a substantial fraction of the MAD-level magnetic flux may still be required to produce a powerful, readily observable jet.

A further constraint is the radial extent of the MRI-active accretion flow ($r_{\rm MRI}$), which must be sufficiently large to supply the magnetic-flux reservoir required for jet launching. Since a detectable jet may require less magnetic flux than MAD formation, we expect
\begin{align}
    r_{\rm MRI} &\gtrsim r_{\rm jet}^{\rm min},\\
    r_{\rm jet}^{\rm min} &\lesssim r_{\rm MAD}^{\rm min},
\end{align}
where $r_{\rm jet}^{\rm min}$ and $r_{\rm MAD}^{\rm min}$ denote the minimum radii enclosing sufficient magnetic flux for observable jet launching and MAD formation, respectively. From Eq.~\ref{eq:rminmad}, we find $r_{\rm MAD}^{\rm min}\simeq110$, $140$, and $1200\,r_g$ for $h/r=0.3$, $0.2$, and $0.1$, respectively. For the thickest disk, this is only a factor of $\sim1.5$ larger than the radius from which the flux responsible for the MAD transition originates, $r\sim80\,r_g$. In contrast, the much larger radius required for the thinnest disk may exceed the extent of some astrophysical accretion flows. Nevertheless, while reaching MAD requires a sufficiently extended disk, observable jets could still be launched from more compact disks with smaller magnetic-flux reservoirs.

Finally, the available magnetic-flux reservoir may evolve as the disk expands through viscous spreading driven by angular-momentum transport. Conversely, turbulent diffusion and reconnection could disrupt large-scale magnetic loops before they reach the black hole, particularly for the thinnest disk, where flux must travel from $\sim10^3\,r_g$. These competing effects introduce additional uncertainties in whether and when the disk can accumulate sufficient flux to reach MAD. We discuss the observational implications in Section~\ref{sec:obs_cons}.

\section{discussion and conclusion}
\subsection{Conclusion}
\label{sec:conclu}

{We have shown that the ability of an accretion disk to generate and accumulate large-scale magnetic flux depends strongly on its thickness. Starting from the same relatively strongly magnetized toroidal field, $\beta_{\rm ini}=5$, disks with smaller $h/r$ develop progressively weaker mean magnetic fields, weaker MRI-driven turbulence, slower inward transport of magnetic flux, and substantially lower electromagnetic jet efficiencies. These effects compound to produce a very steep dependence of the magnetic flux accumulated by the black hole on disk thickness, with the thickest disk reaching the MAD state and jet efficiencies above $100\%$, while the thinnest disk produces only $\sim1\%$ (see Fig.~\ref{fig:phi_BH}). Our results therefore establish a direct physical connection between disk geometry, MRI dynamo action, magnetic-flux transport, and jet launching.}

{MRI turbulence nevertheless generates coherent poloidal magnetic structures in all three simulations despite the absence of an initial net poloidal flux (see Fig.~\ref{fig:field_lines}). In the two thinner disks, these fields organize into coherent large-scale vertical structures that resemble weakly magnetized, magnetically elevated accretion disks \citep{zhu_global_2018,mishra_strongly_2020,jacquemin-ide_magnetic_2021}; see Fig.~\ref{fig:field_lines_avg}. Consistent with this geometry, the radial mass flux is concentrated in layers above and below the midplane, where enhanced inflow velocities compensate for the lower density (see Fig.~\ref{fig:dens_flux}).}

{The strength of both the mean toroidal and poloidal magnetic fields decreases steeply with decreasing $h/r$ (see Fig.~\ref{fig:betas_time}). Together, these results suggest that the weakening of MRI turbulence in thinner disks is closely connected to the reduced strength of the mean poloidal field (see Fig.~\ref{fig:test_law}). In particular, the measured turbulent stresses are reasonably described by established MRI scalings with the vertical magnetic field \citep{hawley_local_1995,salvesen_accretion_2016,lesur_magnetohydrodynamics_2021}. A similar connection between locally generated vertical magnetic flux and turbulent stress has been found in global zero-net-flux simulations \citep{sorathia_connections_2010}. Together, these results suggest that the reduction in turbulent strength is closely linked to the weakening of the mean poloidal field as the disk becomes thinner.}

Our analysis of large-scale magnetic-field generation provides insight into the origin of this dependence. The generation of both large-scale poloidal and toroidal fields becomes increasingly stratified as the disk becomes thinner (see section \ref{sec:field_gen}). In the thinner disks, field generation is concentrated several scale heights above and below the midplane, whereas it is distributed more uniformly throughout the thickest disk (see Fig.~\ref{fig:toroidal_gen}).{ This increasing stratification may be connected to the emergence of the magnetically elevated structure observed in the thinner disks.}

In all simulations, differential rotation is the dominant source of the mean toroidal field, while turbulent terms and axisymmetric induction associated with poloidal flows contribute to its reduction. The latter is dominated by radial advection, with other terms providing smaller contributions to the local balance.

The poloidal field is generated by both non-axisymmetric MRI turbulence and axisymmetric induction, with contributions of comparable importance. The non-axisymmetric contribution is dominated by the largest structures, which appear as coherent rolling motions in the poloidal plane (see Section~\ref{sec:field_gen}, Fig.~\ref{fig:poloidal gen}, and Fig.~\ref{fig:mode_profiles}). Axisymmetric induction also contributes through motions potentially associated with channel-like MRI modes. Since our simulations begin with a purely toroidal field, non-axisymmetric fluctuations are necessary to initially establish the poloidal field, after which both mechanisms contribute to its evolution. The coexistence of these mechanisms is noteworthy, as large-scale non-axisymmetric mean-field generation has primarily been investigated in zero-net-flux shearing boxes \citep{lesur_self-sustained_2008}, while channel-like amplification is typically associated with net-vertical-flux MRI \citep{goodman_parasitic_1994}. However, stratified shearing-box simulations with weak net vertical flux also exhibit large-scale dynamo activity, suggesting that these processes are not mutually exclusive \citep{salvesen_accretion_2016}. Finally, the non-axisymmetric structures become increasingly localized toward the disk atmosphere as $h/r$ decreases (see Fig.~\ref{fig:stream_mode}), while the resulting poloidal structures have a characteristic radial size $l_\varphi\sim h$ (see Appendix~\ref{A:correl_lenght}).

{In contrast, the smallest non-axisymmetric structures predominantly dissipate the poloidal magnetic field, while intermediate-scale structures generate it within the disk but dissipate it in the atmosphere. These regimes are separated by two characteristic wavelengths, \(\lambda_I\) and \(\lambda_S\). When normalized by the disk scale height, both wavelengths are approximately the same across the simulations, and the overall spectrum appears nearly self-similar. This suggests that the characteristic scales separating poloidal-field generation and dissipation are set primarily by the local disk thickness (see section \ref{sec:field_gen} and Fig.~\ref{fig:lenght_scale}).}

The weakening of MRI-driven turbulence also reduces angular-momentum transport and slows accretion (see Figs.~\ref{fig:Stresses} and \ref{fig:accretion_vel}). Remarkably, the inward transport velocity of large-scale magnetic flux follows a similar trend, remaining comparable to the accretion velocity within a factor of $\sim3$. Thus, thinner disks not only generate weaker large-scale magnetic fields but also transport the resulting flux toward the black hole more slowly. These effects combine to produce a steep reduction in the accumulated magnetic flux as $h/r$ decreases. We summarize the measured scalings in Section~\ref{sec:scalings} and show that, taking the scaling of $\beta_p^{-1}$ with $h/r$ as given, the corresponding dependencies of $\alpha$ and $v_{\rm acc}$ follow approximately from simple theoretical arguments.

Using the measured flux distribution and transport velocities, we estimate the timescale required to accumulate MAD-level magnetic flux. This timescale increases rapidly toward smaller $h/r$, reaching $\sim10^8\,r_g/c$ for $h/r=0.1$. Thus, the absence of MAD in the thinner simulations does not necessarily imply that they could never become magnetically arrested. In Section~\ref{sec:obs_cons}, we compare these characteristic timescales with observed accretion transients and discuss their possible relevance for jet launching.
\subsection{Theoretical context and numerical limitations}
\label{sec:context_limits}

A first limitation is numerical resolution. The simulations presented here are among the highest-resolution global calculations of this type and approach the resolution per scale height achieved in some local MRI studies. Nevertheless, whether they fully resolve the asymptotic turbulent state remains uncertain, particularly given the complicated convergence properties of MRI turbulence \citep[e.g.][]{davis_sustained_2010,shi_saturation_2016,ryan_resolution_2017}. As discussed in Appendix~\ref{A:resol}, {we compute the standard MRI quality factors and that they satisfy commonly adopted resolution criteria}. The more stringent mean-field criterion proposed by \citet{jannaud_numerical_2026} suggests that the vertical-field MRI may be marginally resolved in parts of the thinnest disk. However, the measured quality factor ($Q_\theta^{\rm Jann}\sim1.2$) remains within the order-unity uncertainty of the proposed threshold ($Q_\theta^{\rm Jann}\sim2$), making it difficult to draw a definitive conclusion about the degree of numerical resolution. At the same time, several properties of the dynamo appear robust across simulations: in regions where their converged radial domains overlap, the characteristic transition scales $\lambda_I$ and $\lambda_S$ are nearly identical when normalized by $h$, despite the different resolutions per scale height. Higher resolution could still increase the strength of the mean field somewhat, particularly in the thinnest disk, so the precise scalings with $h/r$ discussed in Section~\ref{sec:scalings} should be interpreted with some caution. We nevertheless expect the main qualitative result to be robust: both the mean magnetic field and the associated transport become systematically weaker as the disk becomes thinner, implying a strong dependence on $h/r$, even if the exact steepness of that dependence changes with further increases in resolution.

A second uncertainty is whether the generated magnetic flux will ultimately produce a MAD. For the intermediate-thickness disk, the large flux reservoir and continued inward transport suggest that this may occur if the evolution is continued. For the thinnest disk, however, our estimate requires flux to be supplied from \(r\sim10^3r_g\), making loop survival as important as the nominal advection time. Reconnection, turbulent diffusion, or interactions between neighboring loops could disrupt the field before it reaches the black hole.

A further limitation is that the simulations do not share identical hydrodynamic initial conditions. The two thinner disks begin with a thin-disk density profile, while the thickest disk is initialized from an adiabatic torus. These differences may affect detailed comparisons, particularly the vertical localization of field generation and the less Keplerian accretion profile of the thickest disk. Nevertheless, several properties vary smoothly across the simulations, including the spectral structure of mean-field generation, turbulent-stress profiles, and dominant field-generation mechanisms. We therefore expect the overall trend toward weaker mean fields, turbulence, and flux accumulation in thinner disks to be robust, although the detailed vertical structure and sharpness of the transition with $h/r$ may depend on the initial conditions.

The vertical structure of magnetic-flux transport differs markedly from that found in MAD simulations. We find that flux transport is concentrated in layers above and below the disk midplane, where the radial inflow velocity is largest, so that the magnetic field is effectively carried inward by surface-layer accretion. This behavior is consistent with previous weakly magnetized global simulations in which a magnetically dominated corona efficiently advects large-scale magnetic flux \citep{beckwith_transport_2009,jacquemin-ide_magnetic_2021}, and with theoretical models invoking atmospheric or coronal accretion as a mechanism for flux transport \citep{rothstein_advection_2008,guilet_transport_2012,begelman_simple_2024}. In contrast, \cite{jacquemin-ide_demystifying_2026} find that in MADs both advection and diffusion are concentrated near the midplane, demonstrating that coronal transport is not required once the disk becomes strongly magnetized. Together, these results suggest that the vertical location of magnetic-flux transport depends on the magnetic state of the disk, and potentially also on its thickness and thermodynamics. 

Interestingly, large-scale magnetic-field generation also becomes concentrated in these surface layers as the disk becomes thinner, with both poloidal- and toroidal-field generation moving away from the midplane. The coincidence of field generation and inward transport suggests that the disk atmosphere plays an active role in the large-scale field generation--transport cycle, potentially linking the emergence of a magnetized corona to the vertical localization of MRI activity. We look into this magnetized corona in more detail in a follow-up work \citep{jacquemin-ide_dynamo_2026}.

In a follow-up analysis \citep{jacquemin-ide_dynamo_2026}, we study the dynamo waves propagating through our thinnest disk simulation. We find that an \(\alpha\)-type mean-field prescription provides a useful description only for the thinnest disk, while failing to reproduce the behavior of the two thicker simulations. Using this prescription, we then show that the propagation and structure of the dynamo waves are well reproduced, and examine how these waves contribute to the dissipation of magnetic energy within the corona-like structure of the disk. The success of the \(\alpha\)-type closure only in the thinnest disk may indicate that the MRI dynamo becomes increasingly local as the separation between the turbulent and mean-field scales grows. We discuss this possible connection between scale separation, locality, and the applicability of mean-field closures in detail in that work.

\subsection{Observational consequences}
\label{sec:obs_cons}

The MRI-driven accumulation of magnetic flux considered here may naturally produce two phases of jet activity. Initially, magnetic structures of alternating polarity are advected toward the black hole, producing reversals of the horizon-threading flux. Reconnection between oppositely directed fields can temporarily reduce this flux and weaken the jet, whose power scales approximately with the square of the horizon-threading magnetic flux. Repeated reversals may therefore produce intermittent or striped jets \citep{parfrey_black_2015,christie_role_2019,chashkina_grmhd_2021,kaaz_h-amr_2025,chan_jets_2026}, with current sheets providing sites for reconnection and particle acceleration \citep{lyubarsky_particle_2008,sironi_acceleration_2011}. As progressively larger-scale structures are transported inward, one polarity may eventually dominate, producing a more persistent jet configuration \citep[][see also Fig.~\ref{fig:phi_BH}]{jacquemin-ide_magnetorotational_2024,kaaz_h-amr_2025,chan_jets_2026}.

\emph{Gamma-ray bursts} (GRBs) provide a regime in which magnetic-flux accumulation can occur rapidly. For our thickest disk, $t_{\rm MAD}\sim2\times10^4,r_g/c$, corresponding to only $\sim0.3-0.5$ s for a $3-5,M_\odot$ black hole. Before coherent large-scale flux is established, the accretion of smaller structures with alternating polarity could modulate jet power, potentially contributing to the rapid variability observed in GRBs \citep{golkhou_uncovering_2014}. Both compact-object mergers and collapsars are expected to produce geometrically thick accretion flows, making the $h/r\gtrsim0.2$ regime particularly relevant \citep{siegel_three-dimensional_2017,issa_magnetically_2025,issa_collapsar_2026}. This transition from intermittent to persistent jet activity is also qualitatively reminiscent of precursor emission in short and long GRBs \citep{burlon_precursors_2008,troja_precursors_2010}, although its origin remains uncertain. In neutron-star mergers, precursor emission may instead originate from magnetic activity of a hypermassive neutron-star remnant, which could itself launch a jet before collapsing to a black hole \citep{gottlieb_unified_2025,xiao_peculiar_2024}.

\emph{Black-hole X-ray binaries} provide another possible connection. During the hard-to-soft transition, the inner flow is expected to remain moderately thick, $h/r\sim0.1-0.3$ \citep{done_modelling_2007}, and major discrete relativistic ejections are observed during intermediate states \citep{fender_merlin_1999,fender_jets_2009}. The flux reversals observed in our simulations (see Fig.~\ref{fig:phi_BH}) could contribute to such episodic activity through repeated weakening and re-establishment of the jet. For a $10,M_\odot$ black hole, our estimated accumulation times are $\sim50$ s for $h/r=0.2$ and $\sim1$ hr for $h/r=0.1$, comparable to the shortest recurrence times of repeated ejecta, particularly the tens-of-minutes oscillations of GRS~1915+105 \citep{pooley_variable_1997}. Longer separations of days to tens of days, as observed in XTE~J1859+226 \citep{brocksopp_initial_2002}, likely require additional evolution of the accretion flow. During the reverse soft-to-hard transition, the compact jet re-establishes over days to weeks, with broadband jet emission developing over $\sim10-20$ days in GX~339$-$4 \citep{corbel_formation_2013}. Such timescales could be compatible with slower flux accumulation in disks thinner than $h/r=0.1$, although the evolving disk geometry complicates this comparison.

\emph{AGN} show a less straightforward connection. Radio observations suggest jet episodes lasting $\sim10^7-10^8$ yr, separated by inactive intervals spanning $\sim10^5-10^8$ yr \citep[e.g.][]{shabala_duty_2008,brocksopp_doubledouble_2011,konar_rejuvenated_2012,morganti_archaeology_2017}. For our thinnest disk, the accumulation time of $\sim10^8,r_g/c$ corresponds to only $\sim16-1600$ yr for black-hole masses of $10^6-10^8,M_\odot$, far shorter than typical radio duty cycles. These likely depend on additional processes, such as galaxy-scale fueling or accretion-state changes. Even thinner disks could produce longer accumulation times, although extrapolating our scalings remains uncertain.

\emph{TDEs} exhibit a particularly broad range of outflow-launching timescales. Powerful relativistic jets were already present at the earliest high-energy observations of Swift~J1644+57, Swift~J2058+05, and AT2022cmc \citep{burrows_relativistic_2011,cenko_swift_2012,pasham_birth_2023}. For a $10^6,M_\odot$ black hole, our $h/r=0.3$ model gives an estimated accumulation time of $\sim1$ day, broadly consistent with prompt jet formation. At the other extreme, late-time radio observations infer outflows launched $\sim500-2000$ days after disruption \citep{cendes_mildly_2022,cendes_ubiquitous_2024}. Such delays are also compatible with our scaling: for $M_\bullet\sim10^7,M_\odot$, the $h/r=0.2$ model gives $\sim500$ days, while the $h/r=0.1$ case reaches $\sim15$ yr even for $10^6,M_\odot$. Even if delayed flux accumulation eventually produces a MAD, the accretion rate may have declined by orders of magnitude, resulting in a much weaker jet despite its high efficiency. Magnetic-flux accumulation could thus contribute to both the delayed onset and lower power of late-time outflows, potentially favoring disk thicknesses of $h/r\sim0.1-0.2$ for the delayed population. Delayed circularization and disk formation provide an important alternative explanation \citep{ryu_shocks_2023,chan_jets_2026}.

\emph{Radial extent:} A common uncertainty for transient systems is whether their disks extend far enough to supply the magnetic flux required for MAD formation. Post-merger disks initially form compact tori with outer radii of order tens of $r_g$, but can subsequently expand through viscous spreading driven by angular-momentum transport \citep{fernandez_long-term_2019,de_igniting_2021}. TDE disks may also be compact, with estimates for ASASSN-14li suggesting $R_{\rm out}\sim20-50\,r_g$ from different spectral modeling approaches \citep{wen_opticaluv_2023,guolo_size_2025}. Thus, whether the required flux reservoir can be established at $r^{\rm min}_{\rm MAD}$ depends on the extent and timescale of viscous spreading, rather than the initial disk size alone. However, this radial requirement applies specifically to reaching MAD; launching an observable, albeit weaker, jet may require less magnetic flux and therefore a radial extent smaller by a factor of a few.

Overall, our results provide a physical framework linking disk geometry, MRI dynamo action, magnetic-flux transport, and jet launching. Modest changes in disk thickness can produce orders-of-magnitude differences in the time required to accumulate a jet-capable magnetic field. This may be particularly important for transient accretion systems, where the duration of an accretion episode determines whether sufficient magnetic flux can reach the black hole to launch a powerful jet.

\section*{Acknowledgements}
JJ is indebted to Danat Issa and Beverly Lowell for their help in transferring numerical data. JJ is grateful to François Rincon for his involvement in the early stages of this project. JJ also thanks Geoffroy Lesur, Nick Kaaz, Prasun Dhang, and Stephen Majeski for insightful discussions.

JJ and MCB acknowledge support from NASA grants 80NSSC22K0826, 80NSSC24K0941, and 80NSSC24K1174. J.J. acknowledges support by NSF grants AST-2307983.
AT acknowledges support by NASA
80NSSC26K0343, 
80NSSC22K0031, 
80NSSC22K0799, 
80NSSC18K0565 
and 80NSSC21K1746 
grants, and by the NSF grants 
AST-2009884, 
AST-2107839, 
AST-1815304, 
AST-1911080, 
AST-2206471, 
AST-2407475, 
AST-2510570. 
This research used resources of the National Energy Research
Scientific Computing Center, a DOE Office of Science User Facility
supported by the Office of Science of the U.S. Department of Energy
under Contract No. DE-AC02-05CH11231 using NERSC award
HEP-ERCAP0037682 and NP-ERCAP0029085.  This research used resources of the Oak Ridge Leadership Computing Facility, which is a DOE Office of Science User Facility supported under Contract DE-AC05-00OR22725 via Director Discretionary allocations PHY129.
\section*{Data Availability}

The reduced data products underlying the results presented in this paper are available from the corresponding author upon reasonable request. The full simulation outputs are not publicly archived owing to their size.



\bibliographystyle{mnras}
\bibliography{example} 




\appendix
\section{Radial correlation length scale for the large scale flux}
\label{A:correl_lenght}
\begin{figure}
	\includegraphics[width=\columnwidth]{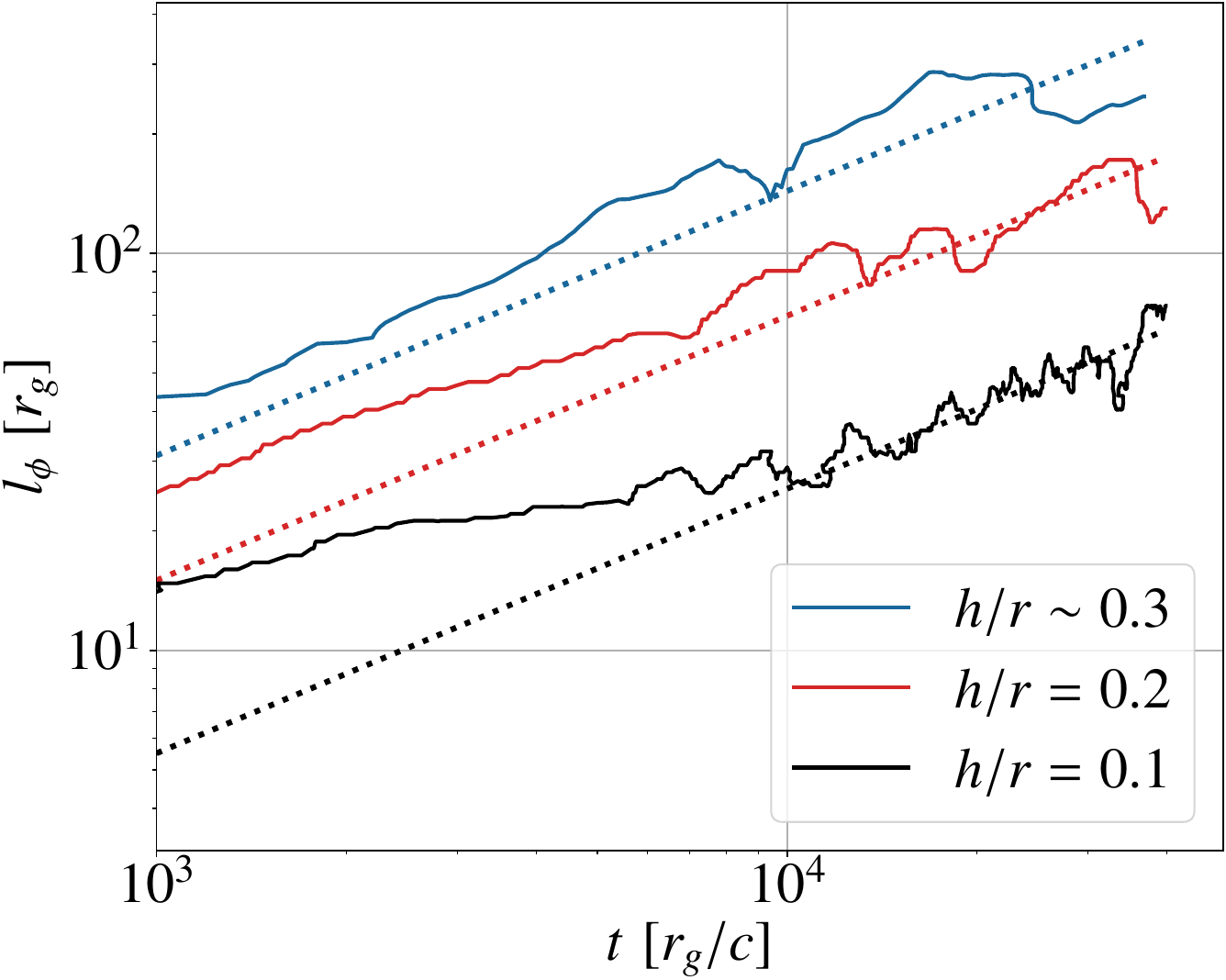}
    \caption{Evolution of the radial magnetic-flux correlation length, $l_\phi$, for the three disk thicknesses. Thicker disks develop systematically larger correlation lengths. The dotted curves show the prediction obtained by assuming that the largest available loops are set by the outermost radius that has completed linear MRI growth and have characteristic size $l_\phi\sim h/f_\phi$. The model closely follows the measured evolution, indicating that the characteristic radial scale of the generated magnetic field is set by the local disk thickness.}
    \label{fig:lenght_scale}
\end{figure}

Another factor contributing to the reduced magnetic flux accumulated onto the black hole is the decrease in the size of the magnetic loops as the disk becomes thinner. Although this trend can be discerned in Fig.~\ref{fig:triple_phi_rt}, it is more clearly quantified by measuring the radial correlation length of the magnetic flux.

To do so, we compute the lagged radial correlation function of $\Phi$,
\begin{equation}
\label{eq:cor}
C_{\rm or}(t,l_{\rm lag}) = \frac{\int\limits^{r_{\rm max}-l_{\rm lag}}_{r_H} \Phi(t,r+l_{\rm lag})\Phi(t,r)\,\mathrm{d}r}{\sqrt{\int\limits^{r_{\rm max}-l_{\rm lag}}_{r_H} \Phi^2(t,r+l_{\rm lag})\,\mathrm{d}r \int\limits^{r_{\rm max}-l_{\rm lag}}_{r_H} \Phi^2(t,r)\,\mathrm{d}r}},
\end{equation}
and define the radial correlation length $l_\phi$ as the first zero of this function,
\begin{equation}
    C(t,l_\phi) = 0.
\end{equation}
Repeating this procedure at every time for all three simulations yields Fig.~\ref{fig:lenght_scale}, which shows the evolution of $l_\phi$ for the different disk thicknesses. As expected, thicker disks exhibit systematically larger magnetic correlation lengths.

The growth of $l_\phi$ can be understood from the outward progression of the MRI. Assuming that the poloidal field grows on approximately the linear MRI timescale,
\begin{equation}
    t_{\rm lin} = \frac{4}{3}\Tfrac{r}{r_g}^{3/2}r_g/c,
\end{equation}
the outermost radius that has completed the linear growth phase evolves as $R_{\rm lin}(t)\propto t^{2/3}$. Since the {largest} characteristic size of a newly generated magnetic loop or {large scale field} structure is approximately the local scale height, 
\begin{equation}
    l_\phi=h(r)/f_\phi
\end{equation}
where $f_\phi$ is a factor of order unity ($0.8$--$1.5$), the largest available magnetic structure at time $t$ is expected to satisfy
\begin{equation}
    l_\phi = \frac{R_{\rm lin}(t)}{f_\phi}\frac{h}{r} = \frac{h/r}{f_\phi}\Tfrac{3tc}{4r_g}^{2/3}r_g.
\end{equation}

This prediction is shown as the dotted curves in Fig.~\ref{fig:lenght_scale}, adopting $f_\phi=0.
8$, $1.1$, and $1.5$ for the $h/r\sim0.3$, $0.2$, and $0.1$ simulations, respectively. The agreement with the measured correlation lengths is remarkably good.

These results suggest that the radial correlation length is set by the largest magnetic loops that have had sufficient time to complete the linear MRI growth phase. Since the size of these loops remains proportional to the local thermal scale height, $l_\phi\propto H$, the characteristic scale of the generated magnetic field decreases rapidly with decreasing $h/r$. This provides another explanation for the steep reduction in the magnetic flux accumulated onto the black hole in thinner accretion disks.
\section{Resolution requirements}
\label{A:resol}
We assess the numerical resolution of the MRI in our simulations using both the standard quality factors commonly adopted in global accretion-disk simulations \citep[e.g.][]{hawley_assessing_2011} and the recently proposed quality factors of \citet{jannaud_numerical_2026}. These quantities measure the number of grid cells resolving the characteristic MRI wavelength,
\begin{equation}
    \frac{\lambda_{\rm MRI}}{h}
    = 2\pi\sqrt{2\beta^{-1}}.
\end{equation}
The main difference between the two definitions is the magnetic field entering the estimate. The standard quality factors are based on the total magnetic energy, and therefore include both the mean field and its fluctuations, whereas the quality factors proposed by \citet{jannaud_numerical_2026} are evaluated using only the mean magnetic field.

For this purpose, we define the inverse plasma betas associated with the total magnetic energy in the $\theta$ and $\varphi$ directions as
\begin{align}
    \beta_{\theta,\,{\rm tot}}^{-1}
    &= \frac{\mean{B_\theta^2}}{8\pi\mean{P_0}},\\
    \beta_{\varphi,\,{\rm tot}}^{-1}
    &= \frac{\mean{B_\varphi^2}}{8\pi\mean{P_0}}.
\end{align}
These should be distinguished from $\beta_\theta^{-1}$ and $\beta_\varphi^{-1}$ used elsewhere in this work, which contain only the mean magnetic field and exclude the contribution from the fluctuating field, $\delta B^2$. We use the total plasma betas to calculate the standard MRI quality factors and the mean-field plasma betas for the \citet{jannaud_numerical_2026} quality factors.

The corresponding vertical quality factors are
\begin{align}
    Q_\theta^{\rm stand}
    &= 2\pi N_h\sqrt{2\beta_{\theta,\,{\rm tot}}^{-1}},\\
    Q_\theta^{\rm Jann}
    &= 2\pi N_h\sqrt{2\beta_\theta^{-1}},
\end{align}
where $N_H$ is the number of grid cells per scale height near the disk midplane. We have $N_h=64$ for $h/r=0.1$, $N_h=128$ for $h/r=0.2$, and $N_h\simeq90$ for $h/r\simeq0.3$ \citep{liska_large-scale_2020}.

Similarly, the standard azimuthal quality factors is
\begin{equation}
    Q_\varphi^{\rm stand}
    = N_\varphi\frac{h}{r}
       \sqrt{2\beta_{\varphi,\,{\rm tot}}^{-1}},
\end{equation}
where $N_\varphi$ is the total number of grid cells in the azimuthal direction. 

\begin{figure}
	\includegraphics[width=\linewidth]{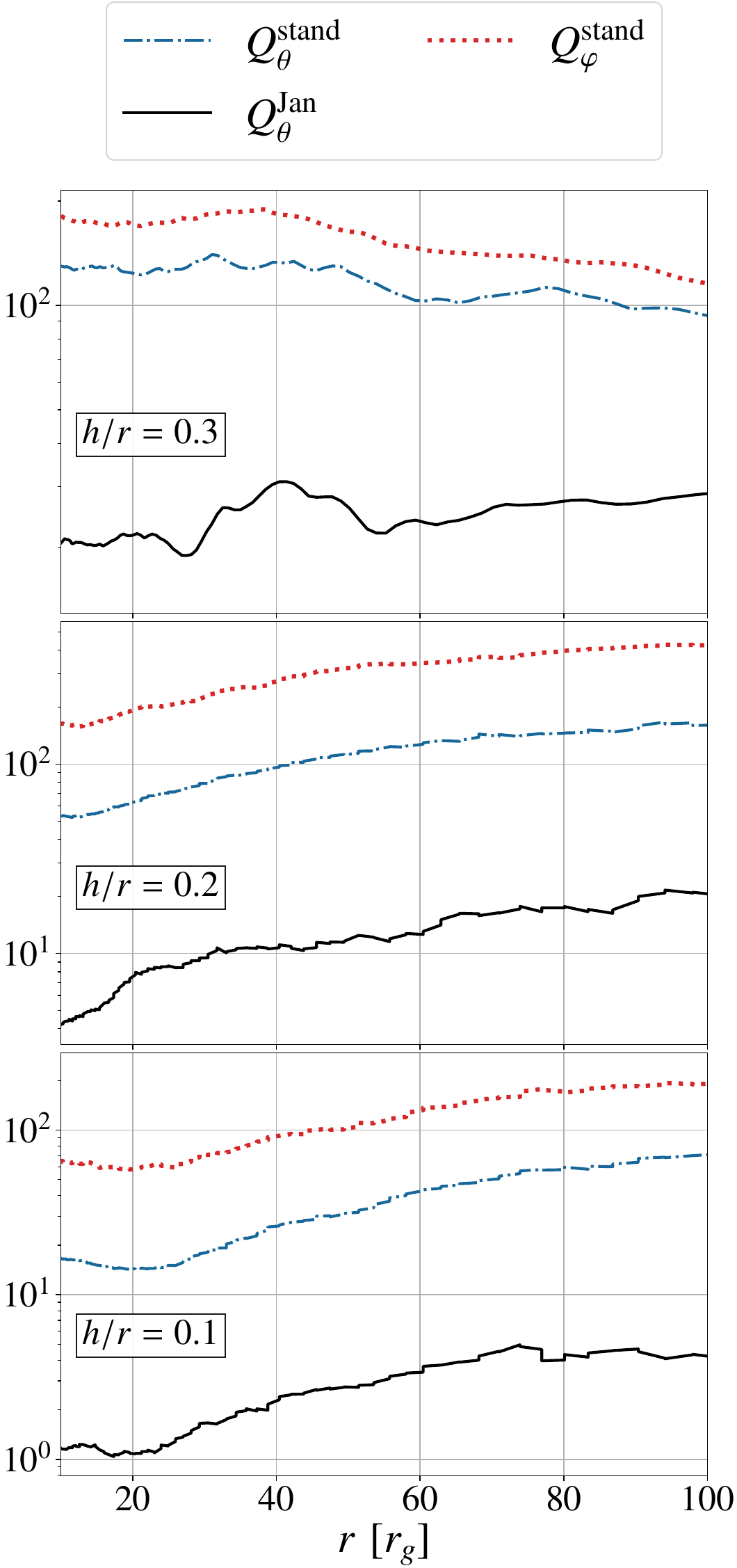}
    \caption{Radial profiles of the standard and mean-field MRI quality factors in the vertical and azimuthal directions for the three disk thicknesses. The standard criteria are satisfied in all simulations, with $Q_\theta^{\rm stand}\gtrsim15$ and $Q_\varphi^{\rm stand}\gtrsim20$. The more stringent mean-field criterion of \citet{jannaud_numerical_2026}  for the vertical field, $Q_\theta^{\rm Jann}$, falls below their proposed threshold of $\sim2$ at $r\lesssim30r_g$ in the $h/r=0.1$ simulation, suggesting some numerical uncertainty in the mean vertical-field MRI. However, \citet{jannaud_numerical_2026} mention order unity uncertainties in their criteria. }
    \label{fig:qual_factors}
\end{figure}

We show these quality factors in Fig.~\ref{fig:qual_factors}. The standard vertical MRI quality factors satisfy $Q_\theta^{\rm stand}\gtrsim15$ in all simulations, consistent with commonly adopted resolution criteria for global MRI simulations \citep{hawley_assessing_2011,hawley_testing_2013}. The standard toroidal quality factors are also well resolved, with $Q_\varphi^{\rm stand}$ comfortably exceeding the commonly adopted threshold of $\sim20$ in all simulations.

{The mean-field quality factors proposed by \citet{jannaud_numerical_2026} provide a more stringent test. According to this criterion, the MRI remains well resolved in the thickest and intermediate-thickness simulations, with $Q_\theta^{\rm Jann}$ comfortably above the proposed threshold of $\sim2$. In the $h/r=0.1$ simulation, however, $Q_\theta^{\rm Jann}$ falls below this threshold at $r\lesssim30r_g$, reaching values as low as $\sim1.2$. }

This possible under-resolution of the mean vertical-field MRI in the thinnest disk introduces an additional numerical uncertainty. However, \cite{jannaud_numerical_2026} note that this threshold is uncertain at the order-unity level, so this shortfall should be interpreted with some caution. We therefore interpret the shortfall in $Q_\theta^{\rm Jann}$ as an indication of possible under-resolution rather than as evidence that the overall dynamics are unconverged.

Several aspects of our results further suggest that the impact of this possible under-resolution may be limited. Several properties of the dynamo appear to be converged across simulations despite their different resolutions per scale height. In particular, the characteristic transition wavelengths $\lambda_I$ and $\lambda_S$, which separate the large-, intermediate-, and small-scale contributions to the dynamo, are nearly identical between simulations when normalized by the disk scale height $h$. This suggests that the scale hierarchy of the dynamo is captured consistently across the different disk thicknesses. The vertical magnetic field is subsequently shaped by the nonlinear dynamo and magnetic-flux transport, rather than evolving solely through the axisymmetric vertical-field MRI. Nevertheless, higher-resolution simulations will be required to determine whether the dynamics change once the \citet{jannaud_numerical_2026} quality factors are fully resolved throughout the disk.

\bsp	
\label{lastpage}
\end{document}